\documentclass[a4paper, 11pt]{article}
\pdfoutput=1
\usepackage{jcappub} 
\usepackage{lineno}
\usepackage{graphicx}
\usepackage{amsmath}
\usepackage{mathtools}
\usepackage{mathrsfs}
\usepackage{microtype}
\usepackage{cancel}
\usepackage{bigints}
\usepackage{bm}
\usepackage[dvipsnames]{xcolor}
\usepackage{soul}
\usepackage{csquotes}
\usepackage{subcaption}
\usepackage{enumitem}
\usepackage{xspace}
\usepackage{booktabs}
\usepackage{siunitx}
\usepackage{mleftright}
\usepackage{ragged2e}
\usepackage{tensor}
\usepackage[normalem]{ulem}
\usepackage{cleveref}
\usepackage{hyperref}
\usepackage{comment}
\makeatletter
\input{aas_macros.sty}
\let\jnl@style=\relax
\makeatother

\DeclareSIUnit{\year}{yr}
\DeclareSIUnit{\pc}{pc}
\DeclareSIUnit{\kpc}{\kilo\pc}
\DeclareSIUnit{\Mpc}{\mega\pc}
\DeclareSIUnit{\cLight}{\text{\ensuremath{c}}}
\DeclareSIUnit{\hHubble}{\text{\ensuremath{h}}}
\DeclareSIUnit{\Msun}{\text{\ensuremath{M_\odot}}}

\newcommand*{\ddelta}{{\mathchar '26\mkern -10mu\delta_{\mathrm{D}}}}
\newcommand{\bx}{{\bm x}}

\newcommand{\bk}{{\bm k}}
\newcommand{\bp}{{\bm p}}
\newcommand{\bq}{{\bm q}}

\newcommand{\bl}{{\bm l}}
\newcommand{\beq}{\begin{equation}\begin{aligned}\relax}
\newcommand{\eeq}{\end{aligned}\end{equation}}

\newcommand*{\ml}{\mleft}
\newcommand*{\mr}{\mright}
\newcommand{\dl}{\mathrm{d}}

{\makeatletter
    \intertext@
    \global\let\nobreakintertext=\intertext
\makeatother}
\newcount\dummycount
\ExplSyntaxOn
\patchcmd{\nobreakintertext}{\penalty}{\dummycount}{}{}
\patchcmd{\nobreakintertext}{\penalty}{\dummycount}{}{}
\ExplSyntaxOff

\newcommand{\ma}[1]{{\textcolor{black}{{#1}}}}

\newcommand{\sd}[1]{{\textcolor{black}{{#1}}}}

\definecolor{green2}{cmyk}{0.27, 0, 1, 0.52}
\hypersetup{
    colorlinks,            
    linkcolor=green2,      
    citecolor=green2,      
    filecolor=magenta,     
    urlcolor=green2        
}

\title{Growth of Structure in Multi-species Wave Dark Matter}
\author[a]{Mustafa A.\ Amin}
\emailAdd{mustafa.a.amin@rice.edu}
\author[b]{, M. Sten Delos}
\emailAdd{mdelos@carnegiescience.edu}

\affiliation[a]{Department of Physics and Astronomy, Rice University, Houston, TX, 77005, U.S.A.}
\affiliation[b]{Carnegie Observatories, 813 Santa Barbara Street, Pasadena, CA 91101, USA}

\abstract{
We explore the growth of structure in multi-species wave (and particle) dark matter. We derive the evolution of the power spectrum of total density contrasts
for an arbitrary number of component species, density fractions, and initial field power spectra. We also derive cross-spectra for density correlations across or within individual species. Our framework includes cold and warm wave dark matter, which can give rise to significant intrinsic Poisson-like density fluctuations along with scale-dependent evolution connected to the free-streaming and Jeans scales. Such dark matter components could be globally or locally misaligned scalar fields as well as multi-component fields with spin $>0$.
The framework also includes cold and warm particle dark matter in the appropriate limits.
}

\begin{document}

\maketitle
\flushbottom

\section{Introduction}
\label{sec:intro}
More than $80\%$ of the non-relativistic matter in our cosmos seems to interact via gravity alone \cite{Cirelli:2024ssz,Marsh:2024ury}. The detailed properties of this dark sector are unknown. The dark sector could be a single bosonic field (perhaps the QCD axion \cite{Preskill:1982cy,Abbott:1982af,Dine:1982ah}) or a multi-species ensemble with bosonic and fermionic fields/particles.  The different species can  convey information about their  initial conditions and microscopic properties through how they gravitationally cluster and the impact of this clustering on visible matter.  

In this work, we provide a framework for calculating the gravitational growth of structure in multi-species dark matter. We focus on bosonic fields with masses $\ll {\rm eV}$ and employ a wave-dynamical calculation.
A companion paper \cite{Amin:2025ayf} explores the growth of structure of multi-species \emph{particle} dark matter, and our earlier works \cite{Amin:2025dtd,Amin:2025sla} studied single-species cases. As with these earlier works, the framework we present here allows us to treat the growth of structure in cases where there is both random motion (or ``warmth'') and intrinsic Poisson-like fluctuations in the dark matter, along with the usual adiabatic perturbations. Contributions from globally misaligned scalars and cold and warm particle dark matter are also naturally included in the appropriate limits. \ma{We provide a code for calculating the density power spectra in the multi-species case at \url{https://github.com/mustafaaamin/warm-and-random.git}.}

The different species can be multiple scalar fields, each with their own particle masses and density fractions, or the species can be different components of a spin $>0$ field \cite{Marsh:2015xka,Hui:2021tkt,OHare:2024nmr,Eberhardt:2025caq,Antypas:2022asj,Adshead:2021kvl,Amin:2022pzv,Marzola:2017lbt}.
Existence of multi-species/component light bosonic fields is expected in many high-energy theories \cite{Svrcek:2006yi,Arvanitaki:2009fg,Cicoli:2023opf,Allahverdi:2014ppa,Alexander:2024nvi,Sheridan:2024vtt,Petrossian-Byrne:2025mto,Baryakhtar:2026oun}. Different fields may be produced through different mechanisms, leading to a diverse set of initial conditions and evolutionary histories. For example, within the ``string axiverse", some fields might be globally misaligned during inflation \cite{Svrcek:2006yi,Arvanitaki:2009fg}, while others are locally misaligned after inflation \cite{Allahverdi:2014ppa, Petrossian-Byrne:2025mto}. Spin-1 dark photon models \cite{Fabbrichesi:2020wbt,Caputo:2021eaa,Antypas:2022asj} provide another example, where polarization states act as gravitationally coupled components (e.g.~\cite{Adshead:2021kvl,Amin:2022pzv}). Furthermore, even in single-field models, a fraction of the energy density may be sequestered into substructures -- such as solitons (e.g~\cite{Khlopov:1985jw,Kolb:1993hw, Chavanis:2011zm,Schive:2014dra,Zhang:2024bjo,Zhou:2024mea}), miniclusters (e.g.~\cite{Kolb:1993zz,Hogan:1988mp,Zurek:2006sy, Eggemeier:2019khm,Vaquero}), or interference granules (e.g.~\cite{Schive:2014dra, Hui:2016ltb}) -- effectively yielding a multi-component system. These examples form the basis of our theoretical motivation to study multi-field scenarios.

In terms of production mechanisms, non-thermal production for light bosonic fields falls broadly into two classes: inflationary and post-inflationary. Inflationary production of minimally coupled light scalars often leads to globally misaligned fields with a dominant zero mode of the field, leading to negligible warmth and Poisson fluctuations (for a review, see \cite{Kolb:2023ydq}). Post-inflationary scenarios, on the other hand, yield particularly rich phenomenology: causality considerations typically lead to significant gradient energy and the absence of a large zero mode.  These features naturally produce a Poisson-like isocurvature component accompanied by free-streaming- and effective-pressure-induced suppression in the matter power spectrum \cite{Zurek:2006sy,McQuinn, Amin:2022nlh,Liu:2024pjg,Ling:2024qfv,Long:2024imw,Liu:2025lts,Amin:2025sla,Harigaya:2025pox}. Beyond scalars, vector dark matter is typically dominated by subhorizon modes regardless of whether it is produced during or after inflation (e.g.~\cite{Graham,Long:2019lwl,Adshead:2023qiw}).
In addition, post-inflationary production mechanisms often lead to formation of topological defects and/or dense substructures such as miniclusters and solitons \cite{Buschmann:2021sdq,Gorghetto,Gorghetto:2024vnp,Saikawa:2024bta, Lee:2024toz,Narita:2025jeg,Hogan:1988mp,Kolb:1993hw,Amin:2019ums, Arvanitaki:2019rax,Zhang:2025lwr}. 

On the observational front, the effects of warmth and Poisson-like isocurvature fluctuations can be accessible even when they are present in a subdominant component. Such isocurvature would seed perturbations in the dominant species, and the resulting spectrum of density variations can be influenced by a variety of scale-dependent effects in either of the dark matter components.
Our framework naturally includes the effects of free streaming, effective pressure support due to random motion, wave-dynamical ``quantum pressure'', and the scale dependence of the intrinsic isocurvature perturbations. All of these effects can combine to yield a wide variety of density power spectra, and we provide several examples.

We provide a detailed derivation of the growth of density perturbations in multi-species dark matter. Our derivation is more cumbersome compared to our recent works \cite{Amin:2025dtd} and \cite{Amin:2025sla}, where we considered a single species. The results and implications, however, can be stated in a relatively straightforward manner. With this in mind, we start by providing a summary of the main results first (section \ref{sec:summaryPS}) and a discussion of relevant physical scales, along with some concrete examples of multi-species power spectra (section \ref{sec:ExamplesScales}). \ma{We provide  details of the numerical algorithm and code in section \ref{sec:Num}}. We end with a Summary and Discussion in section \ref{sec:summary}. A detailed derivation of the main results is provided in Appendix \ref{sec:Derivation}, while Appendix \ref{sec:AdAlt} contains an alternative approach to one part of the derivation.

\section{Model and Main Results}
\label{sec:summaryPS}
We consider $\mathcal{N}$ non-relativistic scalar fields $\varphi^s(t,\bx) = \Re[\Psi^s(t,\bx)e^{-im_st}]/\sqrt{2m_s}$ with masses $m_s$, indexed by $s=1,\hdots, \mathcal{N}$.\footnote{If all of the species have the same mass $m_s=m$, they can also be thought of as the $\mathcal{N}=2\mathfrak{s}+1$ components of a spin-$\mathfrak{s}$ bosonic field.} 
The fields evolve according to the Schr\"{o}dinger–Poisson system of coupled equations:
\beq
    \label{eq:SPscalar_0}
    i \partial_t \psi^s = -\frac{\nabla^2}{2m_s a^2}\psi^s + m_s \Phi \psi^s ,
    \qquad
    \nabla^2\Phi = \frac{4\pi G}{a}\ml(\rho - \bar{\rho}\mr),
\eeq
where $\psi^s = a^{3/2}\Psi^s$ is the normalized field,
\beq
    \rho(t,\bx) = \sum_s m_s|\Psi^s(t,\bx)|a^3 = \sum_s m_s|\psi^s(t,\bx)|^2=\sum_s {\rho}_s
\eeq
is the co-moving density, and $\bar{\rho}$ is its spatial average. Note that $a$ is the cosmic expansion factor (which we normalize to 1 today) and $\nabla^2$ is the Laplacian with respect to comoving position.

Consider the Fourier transforms $\psi_\bq^{s+}$ and $\psi_\bq^{s-}$ of the fields $\psi^s$ and $\psi^{s*}$. We can think of the $\psi_\bp^{s\pm}$ as random variables with \emph{initial} statistics (at the initial time $t_0$)
\beq
    \label{eq:cstat_0}
    &\langle \psi_\bp^{s+}(t_0) \psi_\bq^{s'-}(t_0)\rangle =
    \frac{\bar{\rho}_s}{m_s} f^s_0(p)\ddelta(\bp + \bq)\delta_{ss'} ,
    \quad \text{with} \quad
    \int_\bp f^s_0(p) = 1 ,
\eeq
where we use the notation $\int_\bq \equiv \int \dl\bq/(2\pi)^3$ and $\ddelta(\bk) \equiv (2\pi)^3\delta_{\mathrm{D}}(\bk)$.
The function $f^s_0(p)$ is a normalized version of the power spectrum of the field $\psi^s$:
\beq
    f^s_0(p) = \frac{m_s}{\bar{\rho}_s} P_{\psi^s}(p),
\eeq
and it is also the one-particle phase space distribution function.

To study the evolution of the matter density, it is convenient to define the bilinear \cite{Levkov:2018kau,AminMay:2024}
\beq
    \label{eq:bilinearf_0}
    \hat{f}_\bk^s(\bp) \equiv \frac{m_s}{\bar{\rho}_s}\psi^{s+}_{\bp+\bk/2}\psi^{s-}_{-\bp+\bk/2}\,.
\eeq
In terms of $\hat{f}_\bk^s(\bp)$, the fractional density perturbation $\delta=(\rho-\bar\rho)/\bar\rho$ can be expressed in Fourier space as
\beq
    \label{eq:delta_0}
    \delta_\bk 
    =\sum_{s}\mathfrak{f}_s\delta_\bk^s
    =\sum_{s}\mathfrak{f}_s\int_\bp \hat{f}_\bk^s(\bp),
\eeq
where we define $\mathfrak{f}_s\equiv \bar{\rho}_s/\bar{\rho}$, the mass fraction of the field $s$.
Now the expectation values $\langle \hat{f}^s_{\bk_1}(\bp_1)\hat{f}^{s'}_{\bk_2}(\bp_2)\rangle$ yield the density contrast power spectrum:
\beq
\label{eq:PStotal_0}
    P_\delta(k_1) \ddelta(\bk_1 + \bk_2) =
    \sum_{s,s'}\mathfrak{f}_s\mathfrak{f}_{s'}\int_{\bp_1,\bp_2} \langle\hat{f}^s_{\bk_1}(\bp_1)\hat{f}^{s'}_{\bk_2}(\bp_2)\rangle ,
\eeq
where we assume statistical homogeneity and isotropy.

Our goal is to understand the evolution of $P_\delta(k)$. To this end, we can use the Sch\"{o}dinger-Poisson system \eqref{eq:SPscalar_0} to obtain evolution equations for the $\psi_\bp^{s\pm}$ and hence an evolution equation for $\langle \hat{f}^s_{\bk_1}(\bp_1)\hat{f}^{s'}_{\bk_2}(\bp_2)\rangle$.
In Appendix~\ref{sec:Derivation}, we derive this equation, obtain the solutions, and use them to write down how the matter power spectrum $P_\delta(k)$ evolves over time.

In this derivation, we assume a universe dominated by matter and radiation, with the Hubble rate $H(y)=({k_{\mathrm{eq}}}/{\sqrt{2}a_{\mathrm{eq}}})y^{-2}\sqrt{1+y}$, where $y=a/a_{\rm eq}$.
The scale factor $a$ at matter–radiation equality is $a_{\mathrm{eq}}\approx 1/3388$, and the comoving wavenumber associated with the horizon size at that time is $k_{\mathrm{eq}} = a_{\mathrm{eq}}H(a_{\mathrm{eq}})\approx \SI{0.01}{\per\Mpc}$ \cite{Planck:2018vyg}.  

We summarize the main results below, which we express in terms of the time parameter $y=a/a_\mathrm{eq}$. 
The evolution of the density contrast power spectrum $P_\delta(y,k)$ is determined once we specify the initial field power spectrum $f_0^s(\bp)$ for each species at some early time $y_0\ll 1$ in the radiation era (but after all of the species are non-relativistic and the relevant scales $k$ are subhorizon).\footnote{We require a nonrelativistic subhorizon configuration because the calculation is nonrelativistic. These are not severe restrictions for fields that are supposed to represent dark matter, however. Some  early universe relativistic effects are included in \cite{Ling:2024qfv}.
} 

\subsection{Total Power Spectrum}
\label{sec:totalPS}
The time evolution of the power spectrum of the total density contrast is:
\beq
    \label{eq:MainResultPS}
    P_{\delta}(y,k) &= \underbrace{{P}^{(\mathrm{ad})}_{\delta}(y_0,k) \ml[\mathcal{T}_k^{(\mathrm{ad})}(y,y_0)\mr]^2 \vphantom{\int_{y_0}^y}}_{\text{adiabatic IC + evolution}} \!\! +
    \underbrace{P^{(\mathrm{iso})}_\delta(y_0,k) \ml[\mathcal{T}_k^{(\rm iso)}(y,y_0)\mr]^2}_{\text{isocurvature IC + evolution}},
\eeq
where the adiabatic and isocurvature transfer functions\footnote{In our context, the Poisson contribution is generated after inflation and is isocurvature in nature. It is uncorrelated with the adiabatic initial conditions from inflation.} are given by
\beq \label{eq:MainResultT}
\mathcal{T}_k^{({\rm iso})}(y,y_0) &=
 \ml[1 + 3 \int_{y_0}^y \! \frac{\dl y'}{\sqrt{1+y'}} \mathcal{T}^{(\mathrm{b})}_k(y, y')\mathcal{T}^{(\mathrm{c})}_k(y,y')\mr]^{1/2}\,,\\
\mathcal{T}^{(\mathrm{ad})}_k(y,y_0) &= \mathcal{T}^{(\mathrm{a})}_k(y,y_0) + \frac{1}{2} \frac{\dl\ln(P^{({\mathrm{ad}})}_{\delta}(y_0,k))}{\dl\ln(y_0)} \sqrt{1+y_0} \, \mathcal{T}^{(\mathrm{b})}_k(y,y_0)\,.
\eeq
Here, $y_0\ll 1$ is at an initial ``time" when all wavenumber-$k$ modes of interest are sub-horizon, and the field modes of interest are non-relativistic; the initial conditions (IC) are specified at that time. The isocurvature ``initial condition" is
\beq\label{eq:P_iso_init}
P_\delta^{(\rm iso)}(y_0,k)=\sum_s \mathfrak{f}_s^2 \int_\bp f^s_0(|\bp+\bk/2|)f^s_0(|\bp-\bk/2|),
\eeq
but note that it represents the total time-independent Poisson-like contribution (at sufficiently small $k$); it is not solely an initial condition. The adiabatic IC is 
\beq\label{eq:P_ad_init}
P_\delta^{(\rm ad)}(y_0,k)\approx 36P_{\mathcal{R}}(k)\left[3+\ln\ml(0.15k/k_{\mathrm{eq}}\mr) - \ln\ml(4/y_0\mr)\right]^2,
\eeq
where $k^3/(2\pi^2)P_\mathcal{R}(k)\approx 2\times 10^{-9}$ \cite{Planck:2018jri} is the primordial curvature power spectrum.

The three different 
$\mathcal{T}_k^{(\mathrm{a}, \mathrm{b}, \mathrm{c})}$ in the above expressions are determined by the following Volterra equations:\footnote{$\mathcal{T}_k^{(\mathrm{a,b})}$ describe the evolution of initial bulk perturbations to the density and the velocity divergence, respectively, whereas $\mathcal{T}_k^{(\mathrm{c})}$ is related to the evolution of the Poisson fluctuations.}
\beq
\label{eq:Ty}
    \mathcal{T}^{(i)}_k(y,y') &= {\mathcal{T}}^{\mathrm{fs}\,(i)}_k(y,y')+\frac{3}{2}\int_{y'}^y \frac{\dl y''}{\sqrt{1+y''}}{\mathcal{T}}^{\mathrm{fs}\,(\mathrm{b})}_k(y,y'') {\mathcal{T}}^{(i)}_k(y'',y')\,\quad i=\mathrm{a,b,c}.
\eeq
Solving these Volterra equations requires a specification of the free-streaming kernels, ${\mathcal{T}}^{\mathrm{fs}\,(\mathrm{a,b,c})}_k$, which can be calculated based on the field power spectra $f^s_0(\bp)$ of every species, which is assumed to not evolve beyond red-shifting of momenta. The free-streaming kernels are given by
\beq
\label{eq:yTfsabc}
    {\mathcal{T}}^{\mathrm{fs}\,(\mathrm{a})}_k(y,y') &=  \sum_s \mathfrak{f}_s\cos\left[\left({k}/{k_{{\rm j}\,s}^{\rm eq}}\right)^{\!2}\mathcal{F}(y,y')/\sqrt{2}\right] \int_\bp f_0^s(\bp) \exp\ml[-i\frac{{\bp} \cdot {\bk}}{(k^{\rm eq}_{{\rm j}\,s})^2}
    \sqrt{2}\mathcal{F}(y,y')\mr],\\
    {\mathcal{T}}^{\mathrm{fs}\,(\mathrm{b})}_k(y,y')
    &= \mathcal{F}(y,y')\sum_s \mathfrak{f}_s\,{\rm sinc}\left[\left({k}/{k_{{\rm j}\,s}^{\rm eq}}\right)^{\!2}\mathcal{F}(y,y')/\sqrt{2}\right]\int_\bp f^s_0(\bp) \exp\ml[-i\frac{{\bp} \cdot {\bk}}{(k^{\rm eq}_{{\rm j}\,s})^2}
    \sqrt{2}\mathcal{F}(y,y')\mr],\\
{\mathcal{T}}^{\mathrm{fs}\,(\mathrm{c})}_k(y,y') &= \frac{1}{P_\delta^{(\rm iso)}(y_0,k)}\sum_s\mathfrak{f}^2_s \int_\bp f^s_0(|\bp+\bk/2|)f^s_0(|\bp-\bk/2|)\exp\ml[-i\frac{{\bp} \cdot {\bk}}{(k^{\rm eq}_{{\rm j}\,s})^2}
    \sqrt{2}\mathcal{F}(y,y')\mr].
\eeq
Here, ${\rm sinc}(x)=\sin(x)/x$, the function $\mathcal{F}(y,y') = \ln\ml[(y/y')(1+\sqrt{1+y'})^2 / (1+\sqrt{1+y})^2\mr]$ captures the functional dependence of the comoving distance traveled by a particle during the time interval between $y'$ and $y$, and the parameter
\beq
\label{eq:kjeq}
k_{{\rm j}\,s}^{\rm eq}\equiv a_{\rm eq}\sqrt{H_{\rm eq}m_s}\,,
\eeq
is the  wave-dynamical Jeans or ``fuzzy" Jeans scale for the species $s$ \cite{Hu:2000ke}.

Note that the Volterra equations \eqref{eq:Ty} are the same as those for the single species case in \cite{AminMay:2024}, only with more complicated ``initial'' functions ($\mathcal{T}^{\mathrm{fs}(i)}$) obtained from the weighted sums of the field power spectra of all the species.

\subsection{Inter/Intra-species Power Spectra}
The above result is for the power spectrum of the total density contrast. It is also possible to obtain more detailed information related to different species. The cross power spectrum of $\mathfrak{f}_s\delta_s$ and $\mathfrak{f}_{s'}\delta_{s'}$ is
\beq\label{eq:Pss}
P_\delta^{ss'}(k)
&=
\mathfrak{f}_s \mathfrak{f}_{s'}P_\delta^{(\mathrm{ad})}(y_0,k)\mathcal{T}_k^{(\mathrm{ad})s}(y,y_0)\mathcal{T}_{k}^{(\mathrm{ad})s'}(y,y_0)
\\&\hphantom{=}
+\mathfrak{f}_s\mathfrak{f}_{s'}\int_\bp f_0^s(|\bp+\bk/2|)f_0^{s'}(|\bp-\bk/2|)\delta_{ss'}
\\&\hphantom{=}
+\frac{3}{2}\,\mathfrak{f}_s \mathfrak{f}_{s'}P_\delta^{(\mathrm{iso})}(y_0,k)\int_{y_0}^y \frac{\dl y'}{\sqrt{1+y'}} \left[
\mathcal{T}^{(\mathrm{b})s}_{k}(y,y')\mathcal{T}^{(\mathrm{c})s'}_{k}(y,y')
+(s\leftrightarrow s')\right].
\eeq
Note that by definition $P_{\delta}=\sum_s\sum_{s'}P_\delta^{ss'}$.
The Volterra equations that need to be solved now are coupled across species (note the summation in the last term below):
\beq
\label{eq:Tis}
    \mathcal{T}^{(i)s}_k(y,y') &= {\mathcal{T}}^{\mathrm{fs}\,(i)s}_k(y,y')+\frac{3}{2}\int_{y'}^y \frac{\dl y''}{\sqrt{1+y''}}{\mathcal{T}}^{\mathrm{fs}\,(\mathrm{b})s}_k(y,y'')\sum_{s'} \mathfrak{f}_{s'}{\mathcal{T}}^{(i)s'}_k(y'',y')\,,
\eeq
where $i=\mathrm{a,b,c}$. The adiabatic transfer function $\mathcal{T}^{(\mathrm{ad})s}_k$ for each species is still given by the second line of \eqref{eq:MainResultT}, with $(i)\rightarrow (i)s$ in the superscript. The species-specific free-streaming kernels ${\mathcal{T}}^{\mathrm{fs}\,(i)s}$ are given by \eqref{eq:yTfsabc} with the summation and one factor of $\mathfrak{f}_s$ removed (meaning that ${\mathcal{T}}^{\mathrm{fs}\,(i)}=\sum_s \mathfrak{f}_s{\mathcal{T}}^{\mathrm{fs}\,(i)s}$).

\section{Examples \& Relevant Scales}
\label{sec:ExamplesScales}
\sd{We now discuss general features of how the power spectrum evolves in scenarios with warm wave dark matter. For concreteness, we focus on
the case where the field power spectra are all $f_0^s(p)=(2\pi)^{3/2}k_{*s}^{-3}e^{-p^2/(2k_{*s}^2)}$, but the discussion below can be adapted to other shapes of $f_0^s$ as long as the $p^3 f_0^s(p)$ peak near characteristic scales $k_{*s}$. Each species may have its own characteristic scale $k_{*s}$ and mass fraction $\mathfrak{f}_s$.
For these Maxwellian $f_0^s$, we have}
\beq
\label{eq:examplePSTf}
&P_\delta^{\rm iso}(y_0,k)=\sum_s\mathfrak{f}_s^2\frac{\pi^{3/2}}{k_{*s}^3}e^{-\frac{k^2}{4k_{*s}^2}}\,,\\
&\mathcal{T}^{{\rm fs}(\mathrm{a})}_k(\eta,\eta')=\sum_s\mathfrak{f}_s\cos\left[\left({k}/{k_{{\rm j}\,s}^{\rm eq}}\right)^{\!2}\mathcal{F}(y,y')/\sqrt{2}\right]\exp\ml[{-\alpha_{k\,s}^2\mathcal{F}^2(y,y')}/{2}\mr]\,,\\
&\mathcal{T}^{{\rm fs}(\mathrm{b})}_k(\eta,\eta')=\mathcal{F}(y,y')\sum_s\mathfrak{f}_s\,{\rm sinc}\left[\left({k}/{k_{{\rm j}\,s}^{\rm eq}}\right)^{\!2}\mathcal{F}(y,y')/\sqrt{2}\right]\ml[{-\alpha_{k\,s}^2\mathcal{F}^2(y,y')}/{2}\mr]\,,\\
&\mathcal{T}^{{\rm fs}(\mathrm{c})}_k(\eta,\eta')=\frac{1}{P_{\delta}^{\rm iso}(\eta_0,k)}\sum_s\mathfrak{f}_s^2\frac{\pi^{3/2}}{k_{*s}^3}\ml[{-\alpha_{k\,s}^2\mathcal{F}^2(y,y')}/{4}\mr]\,,
\eeq
where we define
\beq
\alpha_{k\,s}\equiv \sqrt{2}\frac{k_{*s}}{a_{\rm eq}m_s}\frac{k}{k_{\rm eq}}.
\eeq
This dimensionless wavenumber is a useful way to present these results is an approximately parameter-independent way.
\subsection{Relevant Scales}
\label{sec:RelScales}
We define the velocity dispersion for each species:
\beq
\sigma_s(y)&\equiv\frac{1}{y}\frac{1}{a_{\rm eq}m_s}\left[\frac{1}{3}\int_\bp f_0^s(\bp) p^2\right]^{1/2}=\frac{1}{y}\frac{k_{*s}}{a_{\rm eq}m_s}\equiv \frac{1}{y}\sigma_{{\rm eq}\,s}.
\eeq
For each component, there is a co-moving free-streaming scale $k_{{\rm fs}\,s}$ and a (``warm'') Jeans scale $k_{{\rm J}\,s}$:
\beq\label{eq:scales_J_fs}
k_{{\rm fs}\,s}(y)&\equiv \left[\frac{1}{a_{\rm eq}}\int_{y_0}^{y} \frac{dy'}{{y'}^2 H(y')} \sigma_s(y')\right]^{-1}\sim\frac{k_{\rm eq}}{\sigma_{{\rm eq}\,s}\ln(1/\sigma_{{\rm eq}\,s})}\,,\\
k_{{\rm J}\,s}(y)&\equiv y^{-1/2}\frac{\sqrt{4\pi G\bar{\rho}}}{\sigma_s(y)}=\frac{\sqrt{3y}}{2}\frac{k_{\rm eq}}{\sigma_{{\rm eq}\,s}}\,,
\eeq
where $\bar{\rho}$ is the total density of dark matter today, and the expression after the $\sim$ is evaluated for $y\gtrsim 1$. The co-moving free-streaming length $(k_{{\rm fs}\,s})^{-1}$ is the distance traveled by free particles/waves from the time $y_0$ to $y$, so adiabatic perturbations are suppressed for $k\gtrsim k_{{\rm fs}\,s}$. Meanwhile, below the co-moving Jeans length ($k_{{\rm J}\,s})^{-1}$, the species $s$ cannot cluster efficiently during matter domination, so perturbation growth is suppressed for $k\gtrsim k_{{\rm J}\,s}$.

As we already discussed in section~\ref{sec:totalPS}, the arguments of the $\cos$ and $\rm sinc$ functions in \eqref{eq:examplePSTf} define a ``fuzzy'' Jeans scale:
\beq
k_{{\rm j}\,s}(y)\equiv y^{1/4}k_{\rm eq}\sqrt{{m_s}/{H_{\rm eq}}}
\eeq
(different from $k_{{\rm J}\,s}$; note the lowercase `j').
As noted above, this scale is essentially the same as the fuzzy Jeans scale in \cite{Hu:2000ke,Passaglia:2022bcr}, typically relevant for globally misaligned wave dark matter (i.e. for $k_{*s}\rightarrow 0$). Intuitively, $k_{{\rm j}\,s}$ can be interpreted as the scale of ``quantum pressure'' or wave-dynamical support, whereas $k_{{\rm J}\,s}$ is the scale of effective pressure support due to random motion (``warmth''). Adiabatic perturbations and growth of isocurvature perturbations are suppressed for $k\gtrsim k_{{\rm j}\,s}$.

The initial isocurvature PS only extends up to $k\sim k_{*s}$, at which scale the density contrasts $\delta_s$ in the species $s$ are of order unity. This spectrum falls toward 0 for $k\gg k_{*s}$. For $k\ll k_{*s}$, the initial isocurvature PS for each species has an amplitude $2\pi^2/k_{{\rm wn}\,s}^3$, where
\beq
k_{{\rm wn}\,s}\equiv \mathfrak{f}_s^{-\frac{2}{3}}(4\pi)^{1/6}k_{*s}.
\eeq
This scale in turn determines the scale where the initial isocurvature PS for a given species exceeds the total adiabatic one at matter radiation equality via $\Delta^2_\delta(y=1,k)\sim (k/k_{{\rm wn}\,s})^3$. The isocurvature PS begins to exceed the adiabatic PS at $k\sim 10^{-2}k_{{\rm wn}\,s}$ as long as $k\gg k_{\rm eq}$.

\begin{table*}
    \centering
    \caption{Summary of scales relevant to the evolution of perturbations in each species $s$.}
    \label{tab:scales}
    \begin{tabular}{l l}
        \hline
        Scale & Meaning and physical outcome \\
        \hline\hline

        $k_{\mathrm{fs}\,s}$ & Free-streaming scale; adiabatic perturbations are suppressed for $k\gtrsim k_{\mathrm{fs}\,s}$ \\

        $k_{\mathrm{J}\,s}$ & ``Warm'' Jeans scale; perturbation growth is suppressed for $k\gtrsim k_{\mathrm{J}\,s}$ \\

        $k_{\mathrm{j}\,s}$ & ``Fuzzy'' Jeans scale; perturbations are suppressed for $k\gtrsim k_{\mathrm{j}\,s}$ \\

        $k_{*\,s}$ & Scale momentum of $f_0^s$; initial isocurvature power peaks near $k\sim k_{*\,s}$ \\

        $k_{\mathrm{wn}\,s}$ & Isocurvature PS can exceed the standard adiabatic PS when $k\gtrsim 10^{-2}k_{\mathrm{wn}\,s}$ \\

        \hline
    \end{tabular}
\end{table*}

Table~\ref{tab:scales} summarizes the relevant scales and how they practically affect the matter power spectrum. Although there are four suppression scales, we emphasize their different physical influences:
\begin{itemize}
    \item $k_{\mathrm{fs}\,s}$ is the scale on which initial perturbations in the species $s$ are erased over the full course of cosmic history. However, the initial isocurvature spectrum cannot be erased.
    \item $k_{\mathrm{J}\,s}$ and $k_{\mathrm{j}\,s}$ are scales on which perturbations to the species $s$ are \emph{continuously} erased. (However, again, the initial isocurvature spectrum cannot be erased.)
    \item $k_{*\,s}$ is the scale on which the initial isocurvature spectrum of $s$ is cut off.
\end{itemize}
Also, there are useful relations between the different scales:
\beq\label{eq:scale_ratios}
\frac{k_{\mathrm{J}\,s}^\mathrm{eq}}{k_{\mathrm{fs}\,s}^\mathrm{eq}}\sim \mathcal{O}(10),
\qquad
\frac{k_{{\rm J}\,s}}{k_{{\rm j}\,s}}=\frac{\sqrt{3}}{2} \frac{k_{\rm j\,s}}{k_{*s}}.
\eeq
The superscript ``$\mathrm{eq}$'' indicates the value at the time of matter-radiation equality. 
Note that the precise value of $k_{\mathrm{J}\,s}^\mathrm{eq}/k_{\mathrm{fs}\,s}^\mathrm{eq}$ depends on the duration of nonrelativistic free streaming during the radiation-dominated era.
The second relationship in equation~\eqref{eq:scale_ratios} implies that perturbations in the species $s$ can be suppressed in qualitatively different ways depending on which of $k_{\rm j\,s}$ and $k_{*s}$ is larger.
\begin{itemize}
    \item If $k_{*s}\ll k_{{\rm j}\,s}$, then $k_{{\rm j}\,s}\ll k_{{\rm J}\,s}$, implying that ``fuzzy'' Jeans suppression is more important than ``warm'' Jeans suppression of perturbations to the species $s$. Additionally, the species-$s$ contribution to the isocurvature PS cuts off already at $k\sim k_{*s}$, so Jeans suppression of this contribution is not important (since $k_{{\rm J}\,s}\gg k_{{\rm j}\,s}\gg k_{*s}$).\footnote{This behavior is why the summary in Table~\ref{tab:scales} notes that all perturbations (not just growth) are suppressed for $k\gtrsim k_{{\rm j}\,s}$. Although the initial isocurvature PS cannot be erased, there is not a regime where ``fuzzy'' Jeans suppression in species $s$ has a major impact on the contribution of the same species to the isocurvature PS.}
    \item If $k_{{\rm j}\,s}\ll k_{*s}$, then $k_{{\rm J}\,s}\ll k_{{\rm j}\,s}$, and ``warm'' Jeans suppression is more important than ``fuzzy'' Jeans suppression of perturbations to the species $s$. For $k_{{\rm J}\,s}\ll k\ll k_{*s}$, Jeans suppression holds the species-$s$ contribution to the isocurvature PS at its initial value.
\end{itemize}
The relationship between $k_{*s}$ and $k_{{\rm j}\,s}$ can also be relevant for soliton formation. For example, the ordering can affect whether they form at the time $y\sim 1$ or are delayed \cite{Gorghetto:2024vnp,Amin:2025sla}. Also see a discussion of these scales in \cite{Amin:2025sla,Liu:2025lts}.

\subsection{Wave and Particle Limits}
\label{sec:WPLimits}
\paragraph{Globally Misaligned Wave Dark Matter:} In the limit where $k_{*s}\rightarrow 0$ for a given component $s$, then $f^s_0(\bp)\rightarrow \ddelta(\bp)$. The contribution of this component to the isocurvature spectrum vanishes: $k_{*s}^{-3} e^{-k^2/4k_{*s}^2}\rightarrow 0$ for any $k\ne 0$. Furthermore, $\alpha_{k\,s}\rightarrow 0$, so that in \eqref{eq:examplePSTf}, the contribution to $T^{{\rm fs}(\mathrm{a},\mathrm{b})s}$ are simply cosine and sinc functions, without the exponentials. In this limit this component is the globally misaligned scalar. This component is expected to have a suppression in its PS when the arguments of the trigonometric functions are of order unity, that is, for $k\sim k_{{\rm j}\,s}(y)$. On scales $k/k_{{\rm j}\,s}(y)\ll1 $, the $\rm sinc$ and cosine functions approach unity. Density perturbations in this component behave like those in cold dark matter on these scales.

\paragraph{Particle Dark Matter:} Now suppose that $k_{*s}\rightarrow \infty$ while $k_{*s}/(a_{\rm eq}m_s)$ is held fixed (which also implies $k_{{\rm j}\,s}\rightarrow \infty$). Then, for any $k$, the isocurvature contribution vanishes, the trigonometric functions become unity, and $\alpha_{k\,s}$ is finite. This is the particle limit, and $k_{*s}/(a_{\rm eq}m_s)$ is simply the characteristic velocity scale of the particle distribution at matter-radiation equality. Depending on the value of $k_{*s}/(a_{\rm eq}m_s)$, the species can be warm or cold. There is a particle Jeans scale $k_{{\rm J}\,s}$ above which there is no growth of structure in this dark matter component. 

\begin{figure}
    \centering
    \includegraphics[width=0.9\linewidth]{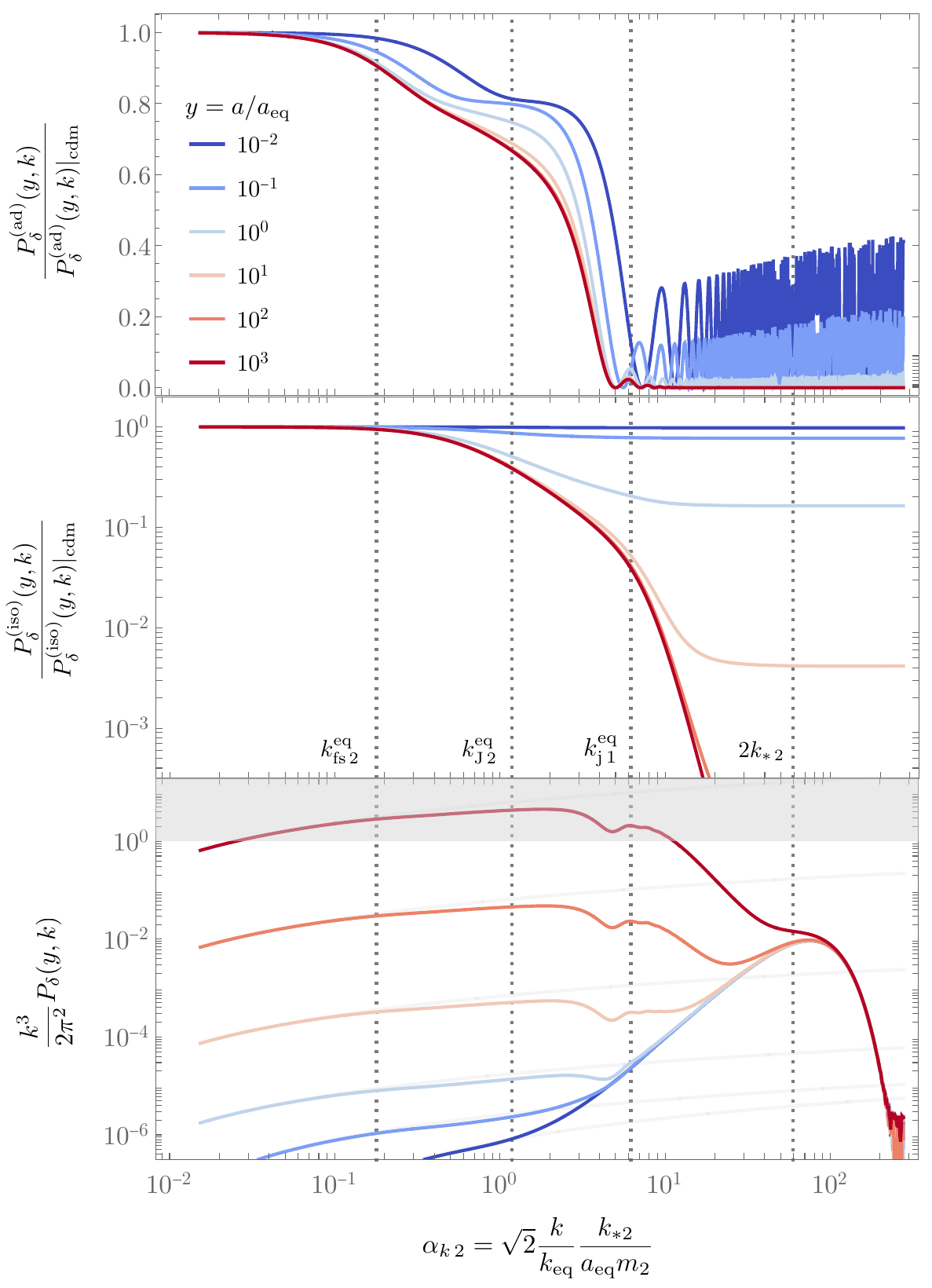}

    \caption{The adiabatic (top panel) and isocurvature (middle panel) contributions to the matter power spectrum for 2-component dark matter relative to single-component CDM. The dominant component 1 has globally misaligned initial conditions for the field, and the subdominant (10\% mass fraction) component 2 has significant Poisson fluctuations and warmth. The bottom panel shows the evolution of the total power spectrum. Here the shaded region delineates nonlinear evolution, and faint gray lines show the behavior of usual CDM with adiabatic ICs. To convert the horizontal axis to a dimensionful $k$, use $k\approx 10^2\,{\rm Mpc}^{-1}\left({22\,{\rm km}\,{\rm s}^{-1}}/{\sigma_{{\rm eq}\,2}}\right)\alpha_{k\,2}$.
    }
    \label{fig:Case1}
\end{figure}

\begin{figure}
    \centering
    \includegraphics[width=0.9\linewidth]{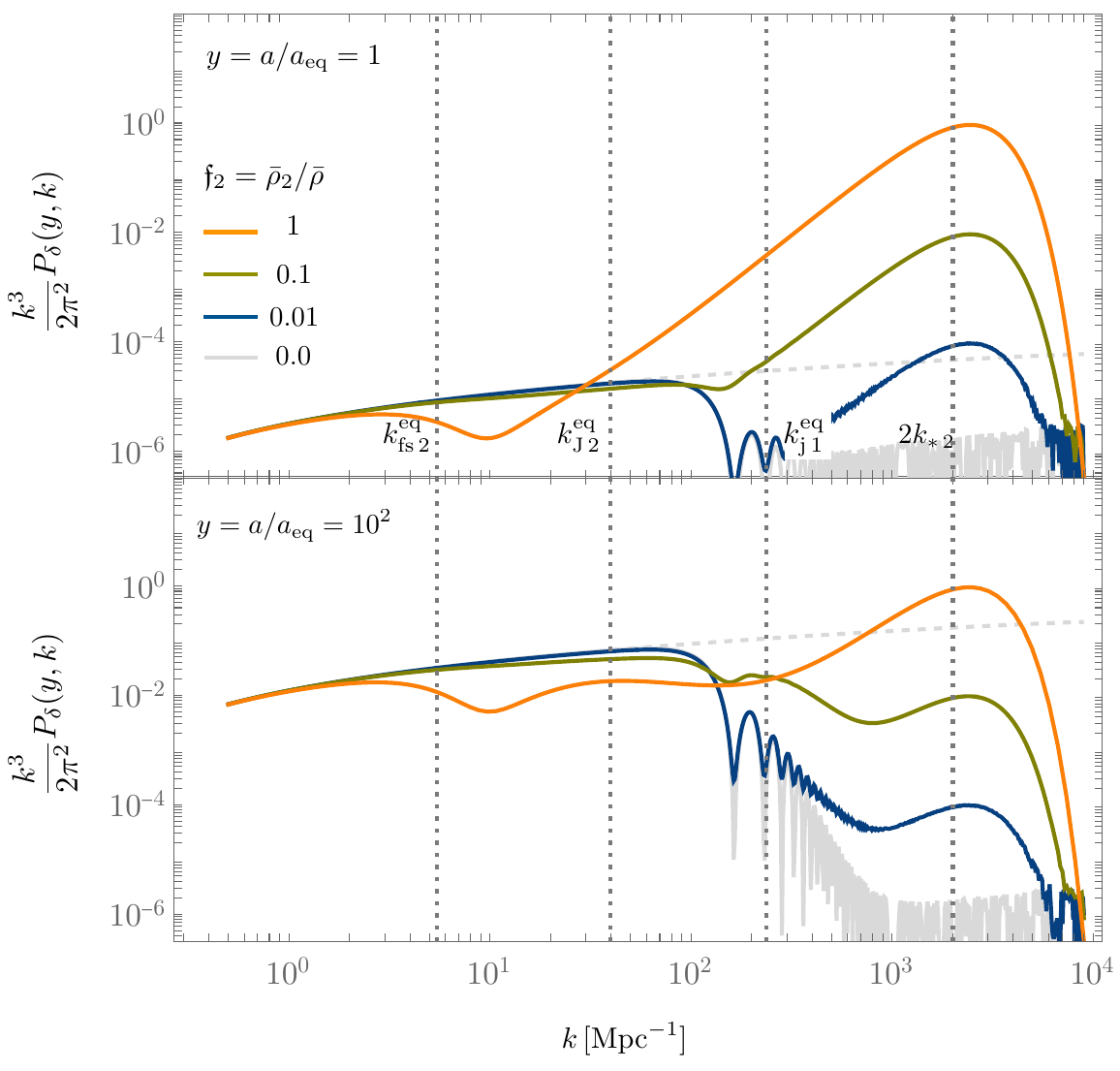}
    \caption{The total power spectrum at $y=1$ (upper panel) and $y=100$ (lower panel), for different density fractions $\mathfrak{f}_2$ of the warm wave dark matter component with $m_2=10^{-19}\,{\rm eV}$ and $k_{*2}=10^3\,{\rm Mpc}^{-1}$ (which yields $\sigma_{\rm eq\,2}\approx 65\,{\rm km}\,s^{-1}$). The other component is a cold, globally misaligned scalar with $m_1=m_2$ and $k_{*1}\rightarrow 0$. Apart from varying $\mathfrak{f}_2$, the scenario is the same as in Fig.~\ref{fig:Case1}. The dashed gray line is CDM.}
    \label{fig:FWCase1}
\end{figure}

\subsection{Examples with a subdominant warm wave dark matter component}

We now present examples of structure formation with a subdominant component of warm wave dark matter. \sd{For simplicity, we restrict our consideration to scenarios with two dark matter components, although the formalism can be applied to arbitrarily many components. For these example scenarios, we} choose parameters so that some of the relevant scales discussed in section \ref{sec:RelScales} lie within observationally accessible regimes.
For observational context, the measurements of the Lyman $\alpha$ forest and high-redshift galaxy luminosity functions constrain the matter power spectrum on scales $k_{\rm obs}\gtrsim 10\,{\rm Mpc}^{-1}$, albeit with order-unity deviations allowed \cite{Chabanier:2019eai,Sabti:2021unj,Garcia-Gallego:2025kiw,Bird:2023evb,Irsic:2019iff,Boylan-Kolchin:2022kae,Irsic:2023equ, Ivanov:2023yla,DESI:2024lzq,Trost:2024ciu,Lazare:2024uvj,He:2025jwp, Buckley:2025zgh}.\footnote{Probes of nonlinear structure 
(e.g.~\cite{Mondino:2020rkn, Gilman:2021gkj, Delos:2021ouc, Hayashi:2021xxu, Delos:2023dwq, Esteban:2023xpk, Nadler:2024ims, Xiao:2024qay, Ji:2024ott, May:2025uldm}) can reach even higher $k$, but their interpretation depends on how the perturbative calculations in this work map to nonlinear structure, a matter we leave for future work. Future high-redshift probes can also reach higher $k$ in the linear regime \cite{deKruijf:2024voc, Munoz:2019hjh}.} Meanwhile, galaxy surveys probe the matter power spectrum at $k_{\rm obs}\sim 0.1$-$1\,{\rm Mpc}^{-1}$ at the level of a few tens of percent \cite{DES:2021wwk,Drlica-Wagner:2022lbd,DESI:2024mwx, Chung:2023syw}.
Heuristically, the parameters $(\mathfrak{f}_s, k_{*s}, \sigma_{{\rm eq}\,s}=k_{*s}/{a_{\rm eq}m})$ of interest observationally are those for which $k_{{\rm fs}\, s}^{\rm eq}\lesssim k_{\rm obs}$   and $\mathfrak{f}_s^2(1/k_{*s})^3\gtrsim P^{(\Lambda {\rm CDM})}_\delta(y,k_{\rm obs})$.\footnote{Assuming a singles species, using a lack of observation of free-streaming suppression or shot noise enhancement in the Ly-$\alpha$ data, a bound on dark matter mass of $m\gtrsim 10^{-19}\,{\rm eV}$ was obtained \cite{Amin:2022nlh} (also see \cite{Long:2024imw}). Using the formalism here, similar techniques can be used to  obtain bounds for subdominant fractions also.} 

We consider two scenarios, each with a dominant cold species with mass fraction $\mathfrak{f}_1=0.9$ and a subdominant warm wave dark matter species with $\mathfrak{f}_2=0.1$. In the first scenario, the dominant species is globally misaligned (cold) wave dark matter with mass $m_1=10^{-19}\,{\rm eV}$, leading to a ``fuzzy'' Jeans scale of $k^{\rm eq}_{{\rm j}\,1}\sim 200\, {\rm Mpc}^{-1}$. In the second scenario, the dominant species is cold particle dark matter with no relevant suppression scale.
In both cases, the subdominant warm wave component has mass $m_2=10^{-19}\,{\rm eV}$ and a field power spectrum $\propto k_{*2}^{-3}e^{-p^2/k_{*2}^2}$ with $k_{*2}=10^{3}\,{\rm Mpc}^{-1}$. These parameters correspond to the velocity dispersion $\sigma_{{\rm eq}\,2}=k_{*2}/(a_{\rm eq}m_2)\approx 65\,{\rm km}\, {\rm s}^{-1}$.
The relevant suppression scales for this component are $k_{{\rm J}\,2}^{\rm eq}\sim 40\,{\rm Mpc}^{-1}$, $k^{\rm eq}_{{\rm fs}\,2}\sim 6\,{\rm Mpc}^{-1}$, and $ k^{\rm eq}_{{\rm j}\,2}\sim 200\, {\rm Mpc}^{-1}$.

\subsubsection{Globally misaligned wave dark matter with a subdominant warm wave component}

For the first example, the dominant dark matter component is a globally misaligned ultralight scalar with $m_1=10^{-19}\,{\rm eV}$ and $k_{*1}\rightarrow 0$. This means that $\sigma_{{\rm eq}\,1}\rightarrow 0$ (the component is cold) and the ``warm'' suppression scales for this component are $k^{\rm eq}_{{\rm J}\,1}\to\infty$ and $k^{\rm eq}_{{\rm fs}\,1}\rightarrow \infty$. The ``fuzzy'' suppression scale for this component is $k^{\rm eq}_{{\rm j}\,1}\sim 200\, {\rm Mpc}^{-1}$, as mentioned above, since it is set solely by the mass. For this scenario, the upper two panels of Fig.~\ref{fig:Case1} depict the evolution of the adiabatic and isocurvature parts of the power spectrum (relative to the evolution of single-component CDM).

In the adiabatic spectrum, there is free-streaming suppression of order $\mathfrak{f}_2=\bar{\rho}_2/\bar{\rho}=0.1$ for $k\gtrsim k_{{\rm fs}\,2}^{\rm eq}$, because free streaming only erases perturbations to component 2. Suppression of the adiabatic spectrum becomes much sharper for $k\gtrsim k_{{\rm j}\,1}^\mathrm{eq}$ due to ``fuzzy'' pressure support in the dominant component 1.
There is also oscillatory power at $k\gtrsim k_{{\rm j}\,1}^\mathrm{eq}$ (coming from $\mathcal{T}^{{\rm fs}(\mathrm{a})}_k$ in equation~\eqref{eq:examplePSTf}), but its lack of growth means that it is already negligible by the onset of the matter dominated era.

Density perturbations in the subdominant component 2 cannot grow for $k\gtrsim k_{{\rm J}\,2}$ due to ``warm'' Jeans suppression, but the initial isocurvature PS of component 2 still seeds perturbations that grow in component 1. Consequently, there is only a gentle suppression of the total isocurvature spectrum for $k_{{\rm J}\,2}^\mathrm{eq}\lesssim k\lesssim k_{{\rm j}\,1}^\mathrm{eq}$. Wave-dynamical (``fuzzy'') pressure support in component 1 leads to much steeper suppression of the isocurvature spectrum for $k\gtrsim k_{{\rm j}\,1}^\mathrm{eq}$.

The bottom panel of Fig~\ref{fig:Case1} shows the total density power spectrum. The light gray curves show the evolution for standard CDM.
Note that the horizontal axis corresponds to $\alpha_{k\,s}= \sqrt{2}\sigma_{{\rm eq}\,2}{k}/{k_{\rm eq}}=\sqrt{3/2}\,k/k_{{\rm J}\,2}^\mathrm{eq}$.

\paragraph{Varying density fractions:} In Fig.~\ref{fig:FWCase1}, we now vary the density fractions $\mathfrak{f}_1$ and $\mathfrak{f}_2$ of components 1 and 2 within the same scenario in which component 1 is a globally misaligned scalar (cold wave dark matter) and component 2 is warm wave dark matter.
By matter-radiation equality ($y=1$), the warm wave dark matter contributes a Poisson-like power spectrum at high $k$, which scales in proportion with $\mathfrak{f}_2^2$.
However, by $y=100$ (redshift $z\sim 30$), the combination of warmth-induced suppression in component 2 with wave-dynamical ``quantum pressure'' support in component 1 leads to a net suppression of the matter power spectrum (in the $\mathfrak{f}_2<1$ cases). This suppression is nevertheless weaker than would arise for single-species cold wave dark matter ($\mathfrak{f}_2=0$).

\subsubsection{Cold particle dark matter with a subdominant warm wave component}
\begin{figure}
    \centering
    \includegraphics[width=0.9\linewidth]{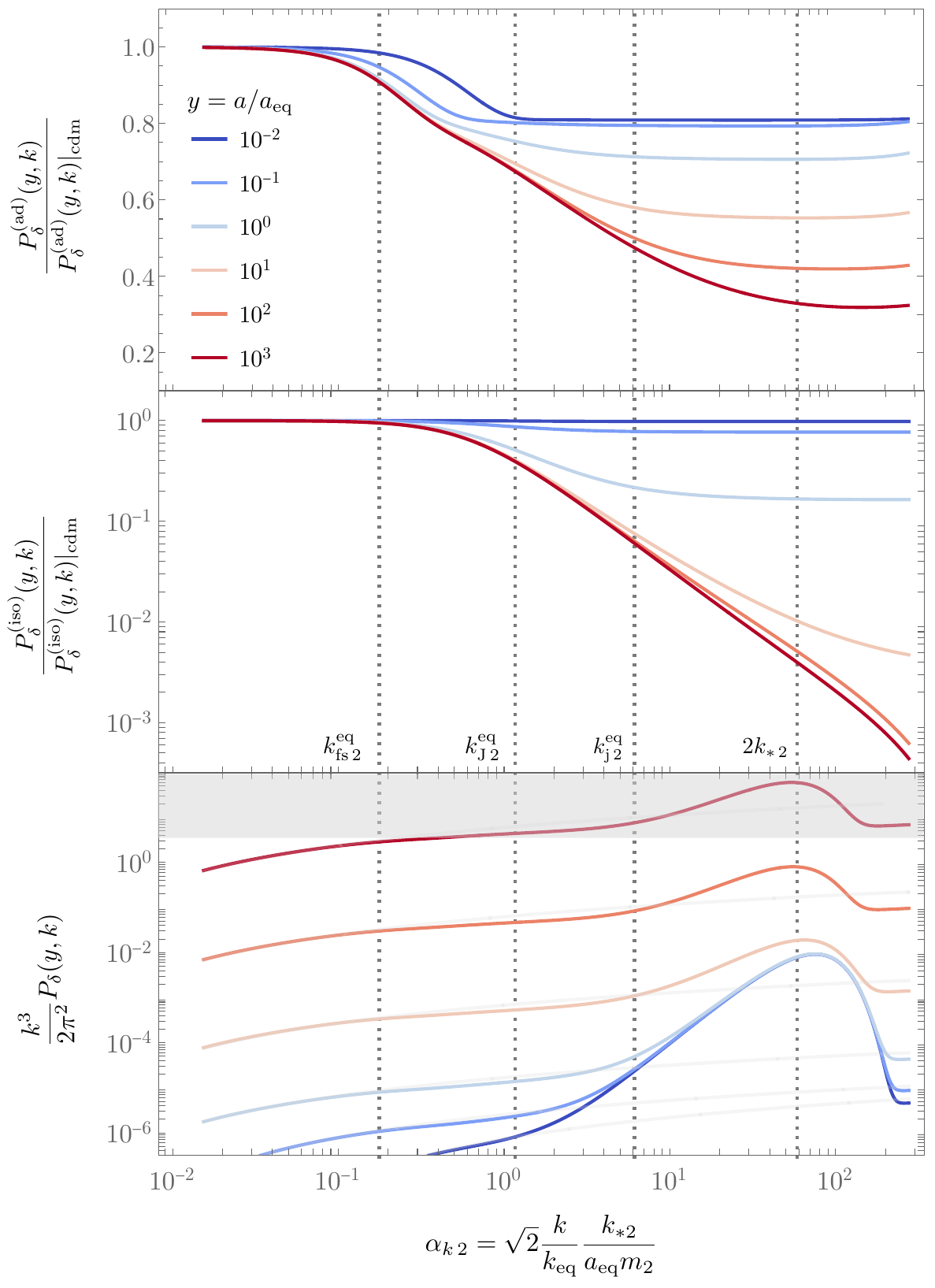}

    \caption{Similar to Fig.~\ref{fig:Case1} but for a scenario where the dominant component 1 is cold and particle-like. As in Fig.~\ref{fig:Case1}, the subdominant component 2 has significant Poisson noise and warmth.
    The upper and middle panels show the adiabatic and isocurvature contributions to the matter power spectrum, respectively, relative to single-component CDM, while the bottom panel shows the total matter power spectrum. In the bottom panel, the shaded region marks where perturbations are nonlinear, and faint gray lines show the behavior of usual CDM with adiabatic ICs. To convert the horizontal axis to wave number, use $k\approx 10^2\,{\rm Mpc}^{-1}\left({22\,{\rm km} s^{-1}}/{\sigma_{{\rm eq}\,s}}\right)\alpha_{k\,s}.$
    }
    \label{fig:Case2}
\end{figure}

For the second example, the dominant dark matter component is cold particle dark matter. Concretely, we set $m_1=10^{-5}\,\rm eV$ and $k_{*1}=10^9\,{\rm Mpc}^{-1}$.
For example, this component could be a QCD-like axion with post-inflationary Peccei–Quinn symmetry breaking (\cite{Gorghetto,Gorghetto:2024vnp,Saikawa:2024bta,Buschmann:2021sdq}).
In this case the velocity dispersion of component 1 is $\sigma_{{\rm eq}\,1}\approx 6.5\times 10^{-7}\,{\rm km}\,{\rm s}^{-1}$, and the suppression scales for this component are $k^{\rm eq}_{{\rm J}\,1}\sim 10^9\,\rm{Mpc}^{-1}$, $k^{\rm eq}_{{\rm fs}\,1}\sim 10^9\,\rm{Mpc}^{-1}$, and $ k^{\rm eq}_{{\rm j}\,1}\sim 10^8\, {\rm Mpc}^{-1}$. These $k$ are too high to affect galaxies or large-scale structure, so this dark matter component is simply CDM.

The evolution of the adiabatic and isocurvature parts of the matter power spectrum (relative to CDM) are provided in the upper two panels of Fig.~\ref{fig:Case2}. The bottom panel of Fig~\ref{fig:Case2} shows the total density power spectrum. Again, the horizontal axis corresponds to $\alpha_{k\,s}= \sqrt{2}\sigma_{{\rm eq}\,2}{k}/{k_{\rm eq}}=\sqrt{3/2}\,k/k_{{\rm J}\,2}^\mathrm{eq}$.

As before, the adiabatic spectrum includes free-streaming suppression of order $\mathfrak{f}_2=0.1$ on scales $k\gtrsim k_{{\rm fs}\,2}^\mathrm{eq}$, since free streaming erases adiabatic perturbations to species 2 only.
On scales $k\gtrsim k_{{\rm J}\,2}^\mathrm{eq}$ that are ``warm"-Jeans suppressed, the adiabatic spectrum is further suppressed after matter-radiation equality because growth of perturbations in component 2 is delayed. When the subdominant component 2 does not participate in gravitational clustering, the growth rate of perturbations to component 1 is lower (e.g.~\cite{Hu:1995en}).

At late times ($y\gg 1$), the isocurvature power spectrum has a gentle $k^{-1}$ suppression for $k\gtrsim k_{{\rm J}\,2}^\mathrm{eq}$.
This scaling arises due to how the initial isocurvature spectrum in component 2, which cannot grow due to the effective pressure support arising from random motion, seeds perturbations to the dominant component 1. High-$k$ isocurvature modes in species 2 have coherence times proportional to $k^{-1}$, so the amplitudes at which they seed power in species 1 are scaled by this factor.

\begin{figure}
    \centering
    \includegraphics[width=0.9\linewidth]{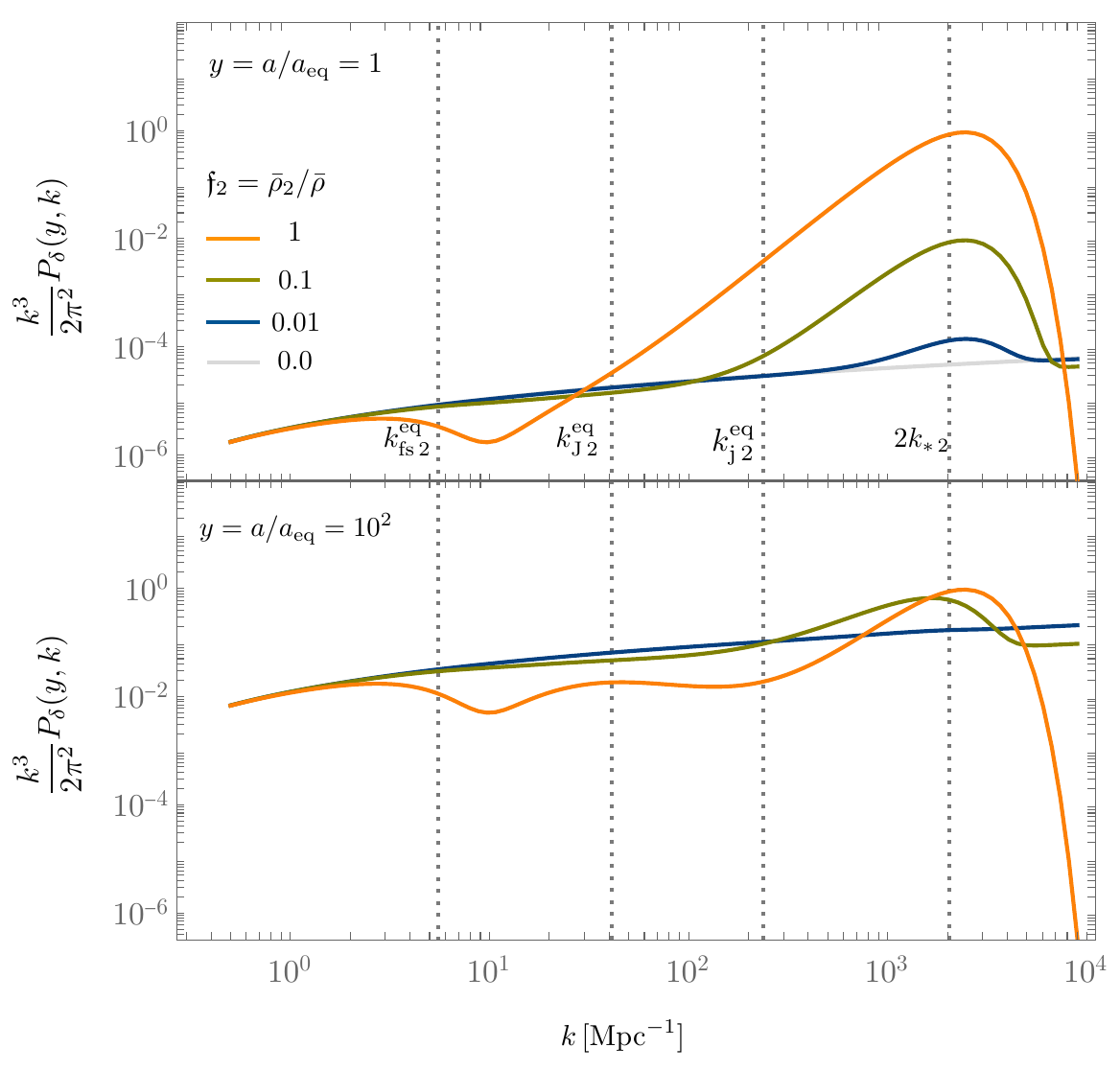}
    \caption{The total power spectrum at $y=1$ (upper panel) and $y=100$ (lower panel), for different density fractions of the warm wave dark matter component. The warm wave component has a mass $m_2=10^{-19}\,{\rm eV}$ and $k_{*2}=10^{3}\,{\rm Mpc}^{-1}$. The other component is effectively CDM with $m_1=10^{-5}\,\rm eV$ and $k_{*1}=10^9\,{\rm Mpc}^{-1}$, which yields $\sigma_{\rm eq 1}\approx 0.0065\,{\rm km}\,s^{-1}$. Apart from varying the fraction $\mathfrak{f}_2$, this is the same case as in Fig.~\ref{fig:Case2}.}
    \label{fig:FWCase2}
\end{figure}

\paragraph{Varying density fractions:} Finally, in Fig.~\ref{fig:FWCase2}, we vary the density fractions $\mathfrak{f}_1$ and $\mathfrak{f}_2$ of components 1 and 2, but we otherwise adopt the same scenario in which component 1 is CDM and component 2 is warm wave dark matter. At the time of matter-radiation equality, the isocurvature contribution from the warm wave dark matter markedly amplifies the matter power spectrum at high $k$ in proportion with $\mathfrak{f}_2^2$. However, by $y=100$ (redshift $z\sim 30$), the warmth of component 2 leads to a much weaker isocurvature rise in the power spectrum. Indeed, the isocurvature rise peaks at a comparable level for $\mathfrak{f}_2=0.1$ as for $\mathfrak{f}_2=1$ at this time because although the initial isocurvature spectrum is lower in the fractional case, it seeds perturbations to the CDM that can then grow without the thermal Jeans suppression. There is also significant suppression of the matter power spectrum at $k\ll k_{*\,2}$ in high-$\mathfrak{f}_2$ cases due to free-streaming and thermal Jeans suppression.

For further discussion of the evolutionary behavior of the power spectrum in multicomponent scenarios, see the perturbative analysis in \cite{Amin:2025ayf}.

\ma{\section{Numerical Algorithm and Code}
\label{sec:Num}
We generalized the iterative algorithm described in \cite{Amin:2025dtd} to obtain the power spectra in the multi-field case.  For the total power spectrum, once $\mathcal{T}^{{\rm fs}\,(\rm i)}_k(y,y')$ have been determined by integrating the initial field spectra $f_0^{s}(p)$, we solve \eqref{eq:Ty} iteratively. At the first iteration, $\mathcal{T}^{{\rm fs}\,(i)}_k(y,y')$ is used as an ansatz. The updated solutions $\mathcal{T}^{(i)}_k(y,y')$ are then substituted back into the right hand side of \eqref{eq:Ty}, and the process is repeated. We find that the solutions converge rapidly: approximately $12$-$13$ iterations are sufficient to achieve percent-level accuracy.}

\ma{The power spectrum evaluation requires calculation of $\mathcal{T}^{(i)}_k(y,y')$ on a grid of $(y,y')$ (with $y>y'$) for each $k$. Combining the iterative $\mathcal{T}^{(i)}_k$ evaluations with the integration scheme for the power spectrum calculation \eqref{eq:MainResultPS}, we find that the results converge as $N_{y}^{-1}$, where $N_{y}$ is the number of log-spaced $\Delta y$ intervals. We have found that $N_{ y}\gtrsim{\rm few}\times 10^2$ leads to reliable power spectra at late times for the examples we considered. Explicitly, for the case in Fig.~\ref{fig:Case2}, with $(y_i,y_{\rm f})=(10^{-3},10^3)$ and $300$ log-spaced $y$-intervals, we get few-percent-level accuracy for $k\lesssim 10^2\,{\rm Mpc}^{-1}$. This accuracy can degrade at higher $k\gg 10^2\,{\rm Mpc}^{-1}$, and it can be improved by increasing $N_{y}$, albeit at increasing computational cost.}

\ma{Our code for calculating the multi-field power spectra can be downloaded at \url{https://github.com/mustafaaamin/warm-and-random.git}. On a modern laptop, the typical examples shown in this paper take between a few seconds and 1 minute to evaluate.}


\section{Discussion \& Summary}
\label{sec:summary}

We have provided a framework for calculating the growth of structure in multi-species dark matter. Starting with the Schr\"{o}dinger–Poisson system for multiple fields, we derived the scale-dependent evolution of the dark matter density power spectrum.
The evolution is described by the solutions to Volterra-type equations, and it is principally controlled by the particle mass $m_s$ of each species and the initial power spectrum of the field, which is peaked near some momentum $k_{*s}$. Our results are general enough to include globally misaligned scalar ``cold wave dark matter'' with $k_{*s}\to 0$, conventional cold and warm particle dark matter models with effectively $k_{*s}\to\infty$ and $m_s\to\infty$ (while their ratio, the velocity dispersion, is held at some finite value), and ``warm wave dark matter'' models with general $k_{*s}$ and $m_s$.

The power spectrum of the dark matter density includes a standard adiabatic part as well as an isocurvature part, which is associated with Poisson-like fluctuations set by the scale $k_{*s}$. The isocurvature contribution can significantly boost the matter power spectrum even if it arises from a subdominant component of the dark matter, since it can seed perturbations in the dominant component. Our calculation also naturally incorporates suppression due to free streaming, effective pressure (Jeans) support due to random motion, and wave-dynamical ``quantum pressure'' support in each of the dark matter components.
The combination of the isocurvature boost to the power spectrum with these different suppressive effects -- all of which can be present to different degrees in different components of the dark matter -- leads to a rich variety of potential signatures in the matter power spectrum.

The generality of our framework allows its applicability to a wide range of scenarios. The dark matter species can be part of an axiverse with local or global misalignment, vector or tensor dark matter, classical warm or cold dark matter, or a mixture of these.
By combining the wave dark matter approach in this work with the particle approach in our companion paper \cite{Amin:2025ayf}, the same cosmological scenario could also include a fraction of massive discrete dark matter systems such as primordial black holes or solitons. \ma{We have provided a publicly available code for evaluating the density power spectra in the multi-species wave and particle dark matter at \url{https://github.com/mustafaaamin/warm-and-random.git}.}

The power spectra that arise from our framework would also influence the properties of cosmic structure in the nonlinear regime, such as the halo mass function and soliton mass functions and formation times \cite{Amin:2025dtd,Amin:2025sla,Gorghetto:2024vnp,Chang:2024fol}. 
These nonlinear effects can be probed by measurements of dark matter substructure \cite{Dai:2019lud, Ramani:2020hdo, Lee:2020wfn, Blinov:2021axd, Lee:2021zqw, Delos:2021rqs, Gilman:2021gkj, Esteban:2023xpk,Delos:2023dwq, Graham:2024hah}. 
However, the same dark matter properties that set the power spectrum would also influence the nonlinear evolution in nontrivial ways. We leave for future work an exploration of nonlinear structure formation in the general multicomponent dark matter scenarios considered in this work (e.g.~\cite{Schwabe:2020,Luu:2024lfq,Eby:2020php,
Luu:2023NestedSolitons,
vanDissel:2024CoreHalo, Jain:2023ojg,
Gosenca:2023MultifieldULDM,
Glennon:2023Simulations,
Guo:2021TwoScalarBEC}).

In the future, it will be useful to compare our formalism with other widely used techniques based on Boltzmann solvers \cite{Ma:1995ey,2022ascl.soft03026G,Blas:2011rf}, or effective field theory approaches \cite{Carrasco:2012cv, Lesgourgues:2011re, Celik:2025wkt,Verdiani:2025jcf}. Including baryonic effects using \cite{AREPO} would be worthwhile to study the  impact on observables \cite{MayAmin:2025Future}. The formalism can also be applied to study  gravitational growth of structure in the very early universe after the end of inflation (e.g.~\cite{Easther:2010mr,Musoke:2019ima}), and can be extended to include the impact of non-gravitational self-interactions (e.g.~\cite{Amin:2019ums,Glennon:2020dxs,Eggemeier:2023nyu,Jain:2022agt,Jain:2023tsr,Mocz:2023adf,Capanelli:2025nrj,Capanelli:2025ykg}). Most importantly, we hope that our work will be useful in looking for hints of or constraining the nature of the potentially complex dark sector using the wealth of current and upcoming observational probes.

\section*{Acknowledgements}
MAA acknowledges helpful discussions  with Sokratis Trifinopoulos, Marco Gorghetto, Georgios Valogiannis, Kimberly Boddy, Giovanni Villadoro, and Mehrdad Mirbabayi.  MAA is supported by DOE grant DE-SC0010103. He gratefully acknowledges the hospitality of ICTP Trieste, where the final stages of this work were completed.

\bibliographystyle{JHEP}
\bibliography{main}

\clearpage
\appendix


\section{Derivation of the Main Results}
\label{sec:Derivation}

In this appendix, we derive the main results shown in section~\ref{sec:summaryPS}. The derivation below is similar to the one presented in \cite{AminMay:2024} which assumed a single species. Here, the derivation is generalized to multiple species.  

\subsection{Notation \& Conventions}
\label{sec:notconv}

We first summarize the conventions and notations that we use in the derivation.
The background Friedmann–Lemaître–Robertson–Walker spacetime is given by
\beq
    \dl s^2 = -\dl t^2 + a^2(t) \dl\bx \cdot \dl\bx \,,
\eeq
where $a(t)$ is the scale factor, $\bx$ are comoving coordinates, and we set $c=1$.
Generally, we use comoving lengths and coordinates unless stated otherwise, e.g. $\bx = a(t)^{-1} \mathbf{r}$, where $\mathbf{r}$ is in proper (``physical") length units. Accordingly, we define $\rho$ as the comoving mass density (mass per comoving volume); $\bar\rho$ is its spatial average and $\rho_s$ is the mass density of species $s$.

Our Fourier conventions are
\beq
    \\
    f(\bx) &= \int \! \frac{\dl\bk}{(2\pi)^3} e^{i\bk \cdot \bx} f(\bk),
    &\!\!
    f(\bk) &= \int \! \dl\bx\, e^{-i\bk \cdot \bx} f(\bx),
    &\!\!
    \int \! \dl\bx\, e^{\pm i(\bk+\bk')\cdot \bx} &= (2\pi)^3\delta_{\mathrm{D}}(\bk+\bk') \,.
\eeq
We also use $f_\bk$ to denote the Fourier transform of $f$. Here $\delta_{\mathrm{D}}$ is the Dirac $\delta$-distribution. We use the following short-hand notation
\beq
    \label{eq:finite_infinite}
    \int_\bx \longleftrightarrow \int \dl \bx \,,\qquad\int_\bk \quad\longleftrightarrow\quad \int \frac{\dl\bk}{(2\pi)^3},\qquad \ddelta(\bk)\longleftrightarrow (2\pi)^3\delta_{\mathrm{D}}(\bk).
\eeq
For a statistically homogeneous field $f$, the power spectrum $P_f$ is defined:
\beq
    \langle f_\bk f_{\bk'}\rangle =(2\pi)^3 P_f(\bk)\delta_{\mathrm{D}}(\bk+\bk') \,,
\eeq
The angled brackets stand for the average over the ensemble. We use $\hbar=c=1$ in the derivation of our results.

\subsection{Evolution equations}

Starting from the setup in section~\ref{sec:summaryPS}, we adopt a convenient time parameter $\eta$, defined by $\dl\eta /\dl t = 1/a^2$. In terms of $\eta$, the Schrödinger–Poisson system becomes:
\beq
    \label{eq:SPscalar}
    i\partial_\eta \psi^s = -\frac{\nabla^2}{2m_s}\psi^s + m_s a^2 \Phi \psi^s,
    \quad
    \nabla^2\Phi = \frac{4\pi G}{a}\ml(\rho - \bar{\rho}\mr),
    \quad
    \rho = \sum_s m_s|\psi^s|^2=\sum_s {\rho}_s.
\eeq
In Fourier space, this system becomes
\beq
    \label{eq:Dcpm}
    &\pm i\partial_\eta \psi^{s\pm}_\bp = \frac{p^2}{2m_s} \psi_\bp^{s\pm} + m_s a^2 \!\int_\bq\!\Phi_\bq \psi^{s\pm}_{\bp-\bq},
    \quad
    \Phi_\bq = -\frac{4\pi G}{aq^2}\rho_\bq ,
    \quad
    \rho_\bq = \sum_s m_s\!\int_\bp\! \psi_\bp^{s+} \psi_{\bq-\bp}^{s-}=\sum_s \rho_\bq^s ,
\eeq
where (as in section~\ref{sec:summaryPS}) $\psi_\bq^{s+}$ and $\psi_\bq^{s-}$ are defined to be the Fourier transforms of $\psi^s$ and $\psi^{s*}$, respectively. The correlation function of the bilinear
$\hat{f}_\bk^s(\bp) \equiv (m_s/\bar{\rho}_s)\psi^{s+}_{\bp+\bk/2}\psi^{s-}_{-\bp+\bk/2}$ discussed in section~\ref{sec:summaryPS} thus evolves according to\footnote{In this appendix, we use $\bar{\rho}_s/\bar{\rho}$ for the density fraction instead of $\mathfrak{f}_s$ to avoid confusion with other quantities which are also represented by the letter $f$.}
\beq
    &i\partial_\eta \langle\hat{f}^s_{\bk_1}(\bp_1)\hat{f}^{s'}_{\bk_2}(\bp_2)\rangle
    \\
    &\quad =\left(\frac{\bp_1 \cdot \bk_1}{m_s}+\frac{\bp_2 \cdot \bk_2}{m_{s'}}\right)\langle\hat{f}^s_{\bk_1}(\bp_1) \hat{f}^{s'}_{\bk_2}(\bp_2)\rangle
    \\
    &\quad \hphantom{=}-\frac{3}{2}\bar{H}_0^2 a\sum_{s''}\frac{\bar{\rho}_{s''}}{\bar{\rho}}\int_{\bq,\bl} \frac{m_s}{q^2} \left\langle\hat{f}^{s''}_\bq(\bl) \left[\hat{f}^s_{\bk_1-\bq}\ml(\bp_1-\bq/2\mr) - \hat{f}^{s}_{\bk_1-\bq}\ml(\bp_1 + \bq/2\mr)\right] \hat{f}^{s'}_{\bk_2}(\bp_2)\right\rangle \\
    &\quad \hphantom{=}-\frac{3}{2}\bar{H}_0^2 a\sum_{s''}\frac{\bar{\rho}_{s''}}{\bar{\rho}}\int_{\bq,\bl} \frac{m_{s'}}{q^2} \left\langle\hat{f}^{s''}_\bq(\bl) \left[\hat{f}^{s'}_{\bk_2-\bq}\ml(\bp_2-\bq/2\mr) - \hat{f}^{s'}_{\bk_2-\bq}\ml(\bp_2 + \bq/2\mr)\right] \hat{f}^{s}_{\bk_1}(\bp_1)\right\rangle.
\eeq
Recall that the density contrast power spectrum is simply the integral over this correlation function:
\beq
\label{eq:PStotal}
    P_\delta(k_1) \ddelta(\bk_1 + \bk_2) =
    \sum_{s,s'}\frac{\bar\rho_s}{\bar\rho}\frac{\bar\rho_{s'}}{\bar\rho}\int_{\bp_1,\bp_2} \langle\hat{f}^s_{\bk_1}(\bp_1)\hat{f}^{s'}_{\bk_2}(\bp_2)\rangle .
\eeq

The correlation function can be written as
\beq
    \langle\hat{f}^s_{\bk_1}(\bp_1)\hat{f}^{s'}_{\bk_2}(\bp_2)\rangle = \underbrace{g^{ss'}_{\bk_1\bk_2}(\bp_1,\bp_2)}_{\text{connected}}+\underbrace{\delta_{ss'}f^s_0(|\bp_1\!+\!\bk_1/2|)f^{s'}_0(|\bp_2\!+\!\bk_2/2|)\ddelta(\bp_1\!-\!\bp_2)\ddelta(\bk_1\!+\!\bk_2)}_{\text{disconnected}}.
\eeq
It has a disconnected (Gaussian) and a connected (non-Gaussian) part. The connected part $g$ can be further split into two parts, one where $g=0$ initially and another where $g\ne0$ initially. In what follows, we will  first obtain the density power spectrum due to the disconnected part. Then we focus on the connected part, $g$, which satisfies equations of the form $\mathcal{D}g=S$ with an initial condition $g=0$. These disconnected and connected parts provide the initial conditions and time-evolution of the isocurvature power spectrum. Finally, we discuss the evolution of the power spectrum in the case where $g\ne0$ initially due to adiabatic perturbations.

\subsection{Isocurvature Perturbations}
\paragraph{Disconnected Contribution:} From statistical homogeneity and isotropy, we can write $g^{ss'}_{\bk_1\bk_2}(\bp_1,\bp_2)=\tilde{g}^{ss'}_{k_1}(\bp_1,\bp_2)\ddelta(\bk_1+\bk_2)$. Then, the equation for $f^s_0(p)$ can be obtained in terms of $\tilde{g}^{ss'}_{k}$ via the $\psi^{s\pm}_\bp$ evolution equations (see \cref{eq:Dcpm}) as
\beq
    \partial_\eta f^s_0(p) 
    &= \frac{3\bar{H}_0^2}{2}a \sum_{s'}\frac{\bar{\rho}_{s'}}{\bar{\rho}}\int_{\bk,\bq} \Delta^s_\bk \tilde{g}^{s's}_k(\bq,\bp),
\eeq
where $\Delta^s_\bk \tilde{g}^{s's}_k(\bq,\bp) = -im_s/k^2 \ml[\tilde{g}^{s's}_k(\bq,\bp+\bk/2)-\tilde{g}^{s's}_k(\bq,\bp-\bk/2)\mr]$. We will ignore this time variation of $f^s_0$ due to $g^{s's}$ when using $f^s_0$ in the equation for $g^{ss'}$ below. This is a good approximation as long as density perturbation evolution is linear. The density power spectrum for this time-independent disconnected part is obtained by integrating $\langle \hat{f}^s_{\bk_1} \hat{f}^{s'}_{\bk_2}\rangle$ over $\bp_1\bp_2$:
\beq
    \label{eq:Piso_initial}
    P_{\delta}(\eta,k) \supset P_{\delta}^{(\mathrm{iso})}(\eta_0,k)=\sum_{s}\left(\frac{\bar{\rho}_s}{\bar{\rho}}\right)^{\!\!2}\int_\bp f^s_0(|\bp-\bk/2|)f^{s}_0(|\bp+\bk/2|),
\eeq
which has an approximately white spectrum at low $k$. That is, if $k_{*s}$ are locations of peaks in $q^3f_0^s(q)$, the $P^{(\rm iso)}_\delta(\eta_0,k)\approx {\rm const.}$ for $k\ll k_{*s}$.

\paragraph{Connected Contribution:} Using the equation for the evolution of the correlation function, as well as the split into connected and disconnected parts, the connected part satisfies an equation of the form $\mathcal{D}g = S$:
\beq
    \label{eq:Dg=S}
    &\partial_\eta g^{ss'}_{\bk_1\bk_2}(\bp_1,\bp_2)+i\left(\frac{\bp_1\cdot\bk_1}{m_s}+\frac{\bp_2\cdot\bk_2}{m_{s'}}\right)g^{ss'}_{\bk_1\bk_2}(\bp_1,\bp_2) - {}
    \\
    &\qquad\frac{3}{2}\bar{H}_0^2a \sum_{s''}\frac{\bar{\rho}_{s''}}{\bar{\rho}}\left[\Delta_{\bk_1}^sf_0^s(\bp_1)\int_\bl g^{s''s'}_{\bk_1\bk_2}(\bl,\bp_2)+\Delta^{s'}_{\bk_2}f^{s'}_0(\bp_2)\int_\bl g^{s''s}_{\bk_2\bk_1}(\bl,\bp_1)\right]\\
    &\approx \frac{3}{2}\bar{H}_0^2a  \left[\frac{\bar{\rho}_{s'}}{\bar{\rho}}\Delta^s_{\bk_1}f^s_0(\bp_1)f^{s'}_0(|\bp_2+\bk_2/2|)f^{s'}_0(|\bp_2-\bk_2/2|)+(1,s\leftrightarrow2,s')\right]\ddelta(\bk_1+\bk_2).
\eeq
Here,  $\Delta^s_\bk f^s_0(p) \equiv (-im_s/k^2) \left[f^s_0(|\bp+\bk/2|)-f^s_0(|\bp-\bk/2|)\right]$. It is worth noting that in the particle case \cite{Amin:2025dtd}, $\Delta^s_\bk f^s_0(p)\rightarrow -im_s\bk/k^2 \cdot \nabla_p f^s_0(p)$. 

The homogeneous equation $\mathcal{D}g=0$ has a solution of the form
\beq
\label{eq:gss'hom}
g^{ss'}_{\bk_1\bk_2}(\bp_1,\bp_2)=\left[\gamma^{(i)s}_{\bk_1}(\bp_1)\gamma^{(j)s'}_{\bk_2}(\bp_2)+i\leftrightarrow j\right]\ddelta(\bk_1+\bk_2),
\eeq
where $\gamma_\bk^{(i)s}$ satisfies 
\beq
\label{eq:gilbert1}
    \left(\partial_\eta + i\frac{\bp\cdot\bk}{m_s} \right)\gamma^{(i)s}_{\bk}(\bp)-\frac{3\bar{H}_0^2a}{2} \Delta^{s}_{\bk} f^{s}_0(\bp)\sum_{s'}\frac{\bar{\rho}_{s'}}{\bar{\rho}}\int_\bl \gamma^{(i)s'}_{\bk}(\bl) = 0 ,
\eeq
with $\bar{H}_0^2 = (8\pi G/3)\bar{\rho}$. 
Equivalently,
\beq
    \label{eq:gilbert}
    \gamma^{(i)s}_{\bk}(\eta,\eta_0,\bp) 
    &=
    \gamma^{(i)s}_\bk (\eta_0, \eta_0,\bp)e^{-i\frac{\bp\cdot \bk}{m}(\eta-\eta_0)}
    \\&\hphantom{=}+
        \frac{3\bar{H}_0^2}{2} \int_{\eta_0}^\eta \dl\eta' a(\eta') e^{-i\frac{\bp\cdot \bk}{m_s}(\eta-\eta')} \Delta^s_{\bk}f^s_0(\bp) \sum_{s'}\frac{\bar{\rho}_{s'}}{\bar{\rho}}\int_\bl \gamma^{(i)s'}_{\bk}(\eta',\eta_0,\bl),
\eeq
which is analogous to the Gilbert equation \cite{Brandenberger:1987} generalized to multiple species.

Let $\gamma^{(\mathrm{b})s}_\bk$ be a solution with the ``initial" condition $\gamma^{(\mathrm{b})s}_\bk(\eta,\eta,\bp)=\Delta^{s}_\bk f^s_0(\bp)$. Similarly, let $\gamma^{(\mathrm{c})s}_\bk$ be a solution with $\gamma^{(\mathrm{c})s}_\bk(\eta,\eta,\bp)= (\bar{\rho}_s/\bar{\rho})f^s_0(|\bp+\bk/2|)f^s_0(|\bp-\bk/2|)$.\footnote{The form of this initial condition can be read off from the right hand side of \eqref{eq:Dg=S}.} Then, the solution to the sourced equation $\mathcal{D}g=S$ can be written as
\beq
g^{ss'}_{\bk_1\bk_2}(\bp_1,\bp_2)=\frac{3}{2}\bar{H}_0^2\int_{\eta_0}^\eta \dl\eta'a(\eta')\left[\gamma_{\bk_1}^{(\mathrm{b})s}(\eta,\eta',\bp_1)\gamma_{\bk_2}^{(\mathrm{c})s'}(\eta,\eta',\bp_2)+(\mathrm{c}\leftrightarrow \mathrm{b})\right]\ddelta(\bk_1+\bk_2).
\eeq
The power spectrum evolution for a given pair of $ss'$ is given by
\beq
\label{eq:Pss'}
P_\delta^{ss'}(\eta,k)\ddelta(\bk_1\!+\!\bk_2)
&\supset \frac{\bar{\rho}_s}{\bar{\rho}}\frac{\bar{\rho}_{s'}}{\bar{\rho}}\int_{\bp_1,\bp_2} g^{ss'}_{\bk_1\bk_2}(\bp_1,\bp_2)\\
&\quad=\frac{3}{2}\bar{H}_0^2 \frac{\bar{\rho}_s}{\rho}\frac{\bar{\rho}_{s'}}{\bar{\rho}}\!\!\int_{\eta_0}^\eta\!\! \dl\eta' a(\eta')\!\left[T_k^{(\mathrm{b})s}(\eta,\eta')T_k^{(\mathrm{c})s'}\!(\eta,\eta')\!+\!(\mathrm{c}\!\leftrightarrow\!\mathrm{b})\right]\ddelta(\bk_1\!+\!\bk_2),
\eeq
where we defined the ``transfer functions" $T_k^{(i)s}$ by integrating the $\gamma^{(i)}_\bk(\bp)$ over momenta:
\beq
    \label{eq:Ti}
    T^{(i)s}_k(\eta,\eta') = \int_\bp \gamma^{(i)s}_\bk(\eta,\eta',\bp).
\eeq
From \eqref{eq:gilbert} and \eqref{eq:Ti}, we see that these $T_k^{(i)s}$ satisfy (with $i=\mathrm{b,c}$)
\beq
\label{eq:Tis-a}
    &T^{(i)s}_{k}(\eta,\eta') =T^{{\rm fs}(i)s}_k(\eta,\eta') +
        \frac{3\bar{H}_0^2}{2} \int_{\eta'}^\eta \dl\eta'' a(\eta') T^{{\rm fs}(\mathrm{b})s}_k(\eta,\eta'') \sum_{s'}\frac{\bar{\rho}_{s'}}{\bar{\rho}}T^{(i)s'}_k(\eta'',\eta'),
\eeq
where $T^{{\rm fs}(\mathrm{b})s}_k(\eta,\eta')$ are free-streaming kernels given by: 
\beq
\label{eq:Tfsbc}
T^{\mathrm{fs}(\mathrm{b})s}_k(\eta,\eta') &= \frac{2m_s}{k^2} \sin\ml[\frac{k^2}{2m_s}(\eta-\eta')\mr]\underbrace{\int_\bp f^s_0(p)e^{-i\frac{\bp\cdot \bk}{m_s}(\eta-\eta')}}_{\equiv T^{\mathrm{fs}\, s}_k},\\
T^{{\rm fs}(\mathrm{c})s}_k(\eta,\eta')&=\frac{\bar{\rho}_s}{\bar{\rho}}\int_\bp f^s_0(|\bp+\bk/2|)f^s_0(|\bp-\bk/2|)e^{-i\frac{\bp\cdot\bk}{m_s}(\eta-\eta')}.
\eeq
Notice that solving \eqref{eq:Tis} requires a solution for {\it all} species $s$ simultaneously. 

It is useful to further define the ``total" transfer functions and free-streaming kernels:
\beq
    \label{eq:SumTfsbc}
    T^{(i)}_k(\eta,\eta') &=\sum_s\frac{\bar{\rho}_s}{\bar{\rho}} T^{({i})s}_k(\eta,\eta') ,
    \\
    T^{\mathrm{fs}\,(\mathrm{b})}_k(\eta,\eta_0) &= \sum_s\frac{\bar{\rho}_s}{\bar{\rho}}T^{\mathrm{fs}(\mathrm{b})s}_k(\eta,\eta'),\\
       T^{\mathrm{fs}\,(\mathrm{c})}_k(\eta,\eta_0) &= \sum_s\frac{\bar{\rho}_s}{\bar{\rho}}T^{\mathrm{fs}(\mathrm{c})s}_k(\eta,\eta'),
\eeq
which, when used in \eqref{eq:Tis}, yields:
\beq
    \label{eq:SumTi}
    T_k^{(i)}(\eta,\eta_0) &= T^{\mathrm{fs}\,(i)}_k(\eta,\eta_0)
    + \frac{3\bar{H}_0^2}{2}\int_{\eta_0}^\eta \dl\eta' a(\eta')T^{\mathrm{fs}\,(\mathrm{b})}_k(\eta,\eta') T_k^{(i)}(\eta',\eta_0)\,.
\eeq
This is identical to the equation if there was a single species, with the understanding that each of the $T$s is a weighted sum over all species. 

Using the total transfer function definitions in \eqref{eq:SumTi} and summing over $s$ and $s'$ in \eqref{eq:Pss'}, we immediately have the evolution of the total isocurvature power:
\beq
\label{eq:PdIso}
P^{(\rm iso)}_\delta(\eta,k)&=
P_{\delta}^{(\rm iso)}(\eta_0,k)+3\bar{H}_0^2\int_{\eta_0}^\eta \dl\eta' a(\eta')\sum_{ss'}\frac{\bar{\rho}_s}{\rho}\frac{\bar{\rho}_{s'}}{\bar{\rho}}T_k^{(\mathrm{b})s}(\eta,\eta')T_k^{(\mathrm{c})s'}(\eta,\eta'),\\
&=P_{\delta}^{(\rm iso)}(\eta_0,k)+3\bar{H}_0^2\int_{\eta_0}^\eta \dl\eta' a(\eta')T_k^{(\mathrm{b})}(\eta,\eta')T_k^{(\mathrm{c})}(\eta,\eta').
\eeq
In the above, we have included the disconnected part, $P_{\delta}^{(\rm iso)}(\eta_0,k)$, which provides a time-independent piece to the isocurvature spectrum.
\subsection{Adiabatic Perturbations}
\label{sec:AdPert}
We now focus on the solution of $\mathcal{D}g=0$ with $g\ne 0$ initially. This solution can still be written in the form \eqref{eq:gss'hom}. Adiabatic initial conditions introduce identical bulk velocity and density perturbations in each component. This can be accounted for via the following ansatz for $g^{ss'}_{\bk_1\bk_2}$:
\beq
\label{eq:gss'ad}
g^{ss'}_{\bk_1\bk_2}(\bp_1,\bp_2)=\gamma^{({\rm ad})s}_{\bk_1}(\bp_1)\gamma^{({\rm ad})s'}_{\bk_2}(\bp_2)\ddelta(\bk_1+\bk_2)
\eeq
with
\beq
\gamma^{({\rm ad})s}_{\bk}(\eta,\eta_0,\bp)
=\sqrt{P^{({\rm ad})}_{\delta}(\eta_0,k)}\gamma_\bk^{(\mathrm{a})s}(\eta,\eta_0,\bp)
+\frac{d\sqrt{P^{({\rm ad})}_{\delta}(\eta_0,k)}}{d\eta_0}\gamma_\bk^{(\mathrm{b})s}(\eta,\eta_0,\bp).
\eeq
The first term captures the density perturbation, and the second is related to the velocity perturbation. For more details, see section 2.2.2 of \cite{Amin:2025dtd}. In the above equation, $\gamma_{\bk}^{(\mathrm{b})s}(\eta,\eta_0,\bp)$ is a solution of \eqref{eq:gilbert} with initial condition $\gamma_{\bk}^{(\mathrm{b})s}(\eta_0,\eta_0,\bp)=\Delta^{s}_\bk f^s_0(\bp)$ (as before), and $\gamma^{(\mathrm{a})s}_\bk(\eta,\eta_0,\bp)$ is the solution with 
\beq
\gamma^{(\mathrm{a})s}_\bk(\eta_0,\eta_0,\bp)=\frac{f^s_0(|\bp-\bk/2)+f^s_0(|\bp+\bk/2)}{2}.
\eeq
We define an additional transfer function $T^{(\mathrm{a})s}_k(\eta,\eta_0)$ via \eqref{eq:Ti} which satisfies \eqref{eq:Tis}, with $T^{{\rm fs}(\mathrm{a})s}_k$  given by 
\beq
    \label{eq:Tfsa}
    T^{\mathrm{fs}(\mathrm{a})s}_k(\eta,\eta_0) &= \cos\ml[\frac{k^2}{2m_s}(\eta-\eta_0)\mr]\int_\bp f^s_0(p)e^{-i\frac{\bp\cdot\bk}{m_s}(\eta-\eta_0)}.
\eeq
The power spectrum, obtained by integrating \eqref{eq:gss'ad} over momenta can be written in terms of these transfer functions is
\beq
P^{({\rm ad})ss'}_\delta(\eta,k)=P_\delta^{({\rm ad})}(\eta_0,k)\frac{\bar{\rho}_s}{\bar{\rho}}\frac{\bar{\rho}_{s'}}{\bar{\rho}}T^{(\mathrm{ad})s}_k(\eta,\eta_0)T^{(\mathrm{ad})s'}_k(\eta,\eta_0),
\eeq
where 
\beq\label{eq:Tad_s}
T^{(\mathrm{ad})s}_k(\eta,\eta_0)=T^{(\mathrm{a})s}_k(\eta,\eta_0)+\frac{1}{2}\frac{d\ln P_\delta^{({\rm ad})}(\eta_0,k)}{d\eta_0}T^{(\mathrm{b})s}_k(\eta,\eta_0).
\eeq
Again, one can define the total $T^{\mathrm{fs}\,(i)}_k$ and
$T^{(i)}_k$ as in \eqref{eq:SumTfsbc}. In terms of these total transfer functions, the total adiabatic power spectrum becomes:
\beq\label{eq:Pad_main}
P^{({\rm ad})}_\delta(\eta,k)=P_\delta^{({\rm ad})}(\eta_0,k)\left[T^{(\mathrm{ad})}_k(\eta,\eta_0)\right]^2,
\eeq
with $T^{(\rm ad)}_k(\eta,\eta_0)=\sum_s (\bar{\rho}_s/\bar{\rho})T^{(\rm ad)s}_k(\eta,\eta_0)$.

We note that in our use of $P_\delta^{(\rm ad)}(\eta_0,k)$ and its derivative, we have first taken an ensemble average over the initial adiabatic perturbations (similar to the approach of \cite{Amin:2025dtd}). An alternative approach would have been to adopt a specific realization of adiabatic perturbations, which would have broken translational symmetry and led to $f_\bk^s(\bp)=\langle\hat{f}^s_{\bk}(\bp)\rangle\ne 0$. We could have then calculated the adiabatic power spectrum using this analogue of the one-particle phase space distribution function. This approach is what was used in \cite{AminMay:2024} for the single species case. For the multicomponent case, we provide this version of the treatment of adiabatic perturbations in Appendix \ref{sec:AdAlt}.
\subsection{Combined Power Spectra}
\label{sec:MainResult}
We arrive at the main result by putting together the adiabatic part (\cref{eq:Tad_s} and \cref{eq:Pad_main}) and the isocurvature part (\cref{eq:PdIso}) to obtain
\beq
    \label{eq:final}
    P_{\delta}(\eta, k)
    = P^{({\rm ad})}(\eta_0,k) \ml[T^{(\mathrm{ad})}_k(\eta,\eta_0)\mr]^2 + P_{\delta}^{(\mathrm{iso})}(\eta_0, k)\ml[T_k^{(\rm iso)}(\eta,\eta_0)\mr]^2,
\eeq
where
\beq
\label{eq:TadTisoSum}
    T_k^{(\mathrm{ad})}(\eta,\eta_0)
    &=T^{(\mathrm{a})}_k(\eta, \eta_0) + \frac{1}{2}\frac{d\ln P_\delta^{({\rm ad})}(\eta_0,k)}{d\eta_0}T^{(\mathrm{b})}_k(\eta, \eta_0),\\
T_k^{(\rm iso)}(\eta,\eta_0)&=\left[1 + 3\bar{H}_0^2\int_{\eta_0}^\eta \dl\eta' a(\eta') T_k^{(\mathrm{b})}(\eta,\eta') \hat{T}_k^{(\mathrm{c})}(\eta,\eta')\right]^{1/2}.
\eeq
Here we define a normalized transfer function $\hat{T}^{(\mathrm{c})}_k = T^{(\mathrm{c})}_k/P_\delta^{(\mathrm{iso})}(\eta_0,k)$. The $T^{(\mathrm{a},\mathrm{b},\mathrm{c})}_k$ can be obtained from \cref{eq:SumTi} with free-streaming kernels in \cref{eq:Tfsbc,eq:Tfsa}.
The expressions above are for the power spectrum of the total density contrast. We can also write down a general expression for the inter- and intra-species power spectra,
\beq
P_\delta^{ss'}(\eta,k)\ddelta(\bk+\bk')&=\frac{\bar{\rho}_s}{\bar{\rho}}\frac{\bar{\rho}_{s'}}{\bar{\rho}}T^{({\rm ad})s}_k(\eta,\eta_0)T^{({\rm ad})s'}_k(\eta,\eta_0)+\delta_{ss'}P^{({\rm iso})s}_\delta(\eta_0,k)\\
&\hphantom{=}+3\bar{H}_0^2\frac{\bar{\rho}_s}{\rho}\frac{\bar{\rho}_{s'}}{\bar{\rho}}\int_{\eta_0}^\eta \dl\eta' a(\eta')T_k^{(\mathrm{b})s}(\eta,\eta')T_k^{(\mathrm{c})s'}(\eta,\eta'),
\eeq
where $T^{(i)s}_k$ are obtained from \eqref{eq:Tis}. Note that $T^{({\rm ad})s}$ has the same form as $T^{({\rm ad})}$ in \eqref{eq:TadTisoSum}, with only the replacement $(i)\rightarrow (i)s$.

For practical calculations, it is convenient to switch to the time parameter $y=a/a_\mathrm{eq}$. To express these results in terms of $y$, we use 
\beq
\dl \eta=\frac{\sqrt{2}}{a_{\rm eq}k_{\rm eq}}\frac{\dl y}{y\sqrt{1+y}},\quad \eta-\eta'=\frac{\sqrt{2}}{a_{\rm eq}k_{\rm eq}}\mathcal{F}(y,y'),\quad y=a(\eta)/a_{\rm eq}.
\eeq
The results are summarized in section \ref{sec:summaryPS}.
\section{Alternate Derivation of the Adiabatic Power Spectrum}
\label{sec:AdAlt}
In section~\ref{sec:AdPert}, we derived the evolution of the adiabatic part of the power spectrum by considering an ensemble of adiabatic realizations as in \cite{Amin:2025dtd}. Here we derive the same evolution considering a specific realization of the adiabatic initial conditions. The results are equivalent.

Given a realization of adiabatic perturbations, the perturbed Fourier coefficients of the field can be computed by a matching procedure.
The density and velocity perturbations $\delta_{\mathrm{ad}}(\eta_0, \bx)$ and $\theta_{\mathrm{ad}}(\eta_0, \bx)$ shift the initial field $\psi^s$ (generated in the absence of adiabatic perturbations $\mathcal{R}$) as follows:
\beq
    \label{eq:psi-ic-a}
    \psi^s(\eta_0, \bx) \rightarrow \psi^s(\eta_0, \bx) \sqrt{1 + \delta_{\mathrm{ad}}(\eta_0, \bx)} \, e^{i S_{\mathrm{ad}}(\eta_0, \bx)} ,
\eeq
with $\nabla^2 S_{\mathrm{ad}} = m\theta_{\mathrm{ad}}$.
At an early enough time $\eta_0$ before free-streaming effects are important but after the dark matter can be treated as a non-relativistic field that interacts only gravitationally, a linear approximation in the density and velocity perturbations yields \cite{Amin:2022nlh}
\beq
    \label{eq:cin-a}
    \psi^{s\pm}_\bp(\eta_0) \to \psi^{s\pm}_\bp(\eta_0) + \frac{1}{2} \int_\bk \delta_\bk^{\mathrm{ad}}(\eta_0) \psi^{s\pm}_{\bp-\bk} \mp \int_\bk\theta_\bk^{\mathrm{ad}}(\eta_0) \frac{im_s}{k^2} \psi^{s\pm}_{\bp-\bk}.
\eeq
Deep in the radiation era, $\delta^{\mathrm{ad}}_\bk(\eta_0) \approx 6\mathcal{R}_\bk (a_{\mathrm{eq}} k_{\mathrm{eq}}/\sqrt{2})(\eta_0 - \eta_k)$ and $\theta_\bk^{\mathrm{ad}}(\eta_0) = -\partial_{\eta_0} \delta^{\mathrm{ad}}_\bk(\eta_0)$, where $\mathcal{R}_\bk $ is the standard adiabatic mode and $\eta_k$ is the time at which $k \approx 1.6a H$ \cite{Dodelson:2020bqr, Baumann:2022mni}. These are good approximations for $k \gg k_{\mathrm{eq}}$. 

After the shift in \cref{eq:cin-a}, $f^s_{\bk}(\bp) \equiv \big\langle\hat f^s_\bk(\bp)\big\rangle \neq 0$ but is proportional to $\delta_\bk^{\mathrm{ad}}(\eta_0)$ and $\theta_\bk^{\mathrm{ad}}(\eta_0)$, which we temporarily treat as independent (cf.~\cref{eq:Pad-a}). The average ``$\langle\hdots\rangle$" here is over initial field configurations. The time evolution of $f^s$ can be written as:
\beq
\label{eq:fsk}
    f^s_{\bk}(\eta,\bp) = \delta_\bk^{\mathrm{ad}}(\eta_0) \gamma^{(\mathrm{a})s}_\bk(\eta, \eta_0, \bp) - \theta_\bk^{\mathrm{ad}}(\eta_0)\gamma_\bk^{(\mathrm{b})s}(\eta,\eta_0,\bp),
\eeq
with the initial conditions
\beq
    \label{eq:gamma-ic-ab-1}
    &\gamma^{(\mathrm{a})s}_\bk (\eta_0, \eta_0, \bp) = \frac{f^s_0(|\bp-\bk/2|)+f^s_0(|\bp+\bk/2|)}{2} ,
    \\
    &\gamma^{(\mathrm{b})s}_\bk (\eta_0, \eta_0, \bp) = \Delta^s_\bk f^s_0(p) \equiv (-im_s/k^2) \left[f^s_0(|\bp+\bk/2|)-f^s_0(|\bp-\bk/2|)\right].
\eeq
We arrive at these by first substituting \cref{eq:cin-a} in $({m_s}/{\bar{\rho}_s})\langle \psi^{s+}_{\bp+\bk/2}(\eta_0)\psi^{s-}_{-\bp+\bk/2}(\eta_0)\rangle\equiv\langle \hat{f}_\bk^s(\eta_0,\bp)\rangle\equiv f^s_\bk(\eta_0,\bp)$, and using \cref{eq:cstat_0} which defines $f_0^s$. This expression for $f^s_\bk(\eta_0,\bp)$ is then substituted on the left-hand side of \eqref{eq:fsk} (evaluated at $\eta=\eta_0$) to read off $\gamma_\bk^{(\mathrm{a,b})s}$ by comparing to the right-hand side of the same equation.

We can use \cref{eq:Dcpm} to derive an equation for $\partial_\eta \hat{f}_\bk^s$. Taking its ensemble average over initial field configurations and linearizing in $\delta_\bk^{\mathrm{ad}}$ and $\theta_\bk^{\mathrm{ad}}$ gives the evolution equation for $\gamma^{s(\mathrm{a}, \mathrm{b})}_\bk$:
\beq
\label{eq:gilbert1-a}
    \left(\partial_\eta + i\frac{\bp\cdot\bk}{m_s} \right)\gamma^{(i)s}_{\bk}(\bp)-\frac{3\bar{H}_0^2a}{2} \Delta^{s}_{\bk} f^{s}_0(\bp)\sum_{s'}\frac{\bar{\rho}_{s'}}{\bar{\rho}}\int_\bl \gamma^{(i)s'}_{\bk}(\bl) \approx 0,
\eeq
where $\bar{H}_0^2 = (8\pi G/3)\bar{\rho}a^3$ and we used $\langle \hat{f} \hat{f}\rangle\approx\langle \hat{f}\rangle \langle\hat{f}\rangle $  in the last term of the equation (hence the approximate sign). This equation is identical to our \eqref{eq:gilbert1}. Integrating the formal solution \eqref{eq:gilbert} over momenta allows us to obtain the transfer functions $T^{(i)s}_k$ (see eq.~\eqref{eq:Tis}). These transfer functions satisfy \eqref{eq:Tis-a} with free-streaming kernels given by \eqref{eq:Tfsbc} for $i=\mathrm{b,c}$ and \eqref{eq:Tfsa} for $i=\mathrm{a}$.

Using the total transfer function and free-streaming kernel definitions as $\bar{\rho}_s/\bar{\rho}$-weighted averages of each species, we obtain formal solutions for $T_k^{(\mathrm{a,b})}$. These can then be used to evolve adiabatic density perturbations:
\beq
\delta_\bk(\eta)
=\delta_\bk^{\mathrm{ad}}(\eta_0)\left[T_k^{(\mathrm{a})}(\eta,\eta_0)+\partial_{\eta_0}\ln\delta_\bk^{\mathrm{ad}}(\eta_0)T_k^{(\mathrm{b})}(\eta,\eta_0)\right]\,.
\eeq
Note that the above equation is just the momentum integral of \eqref{eq:fsk} summed over species, and we have used $\theta_\bk^{(\mathrm{ad})}=-\partial_{\eta_0}\delta_\bk(\eta_0)$. For adiabatic perturbations, $\delta^{\rm ad}_\bk(\eta_0)$ and $\theta_\bk^{(\mathrm{ad})}$ are proportional to $\mathcal{R}_\bk$. Taking the ensemble average over realizations of $\mathcal{R}_\bk$, we can calculate $P^{(\mathrm{ad})}_\delta(\eta,k)\ddelta(\bk+\bk')=\langle \delta_\bk(\eta)\delta_{\bk'}(\eta)\rangle$  to arrive at
\beq
    \label{eq:Pad-a}
    P^{(\rm ad)}_{\delta}(\eta,k) =  P_{\delta}(\eta_0,k) \left[T^{(\mathrm{ad})}_k(\eta,\eta_0)\right]^2,\quad
\eeq
with 
 \beq
 T^{(\mathrm{ad})}_k(\eta,\eta_0)=T^{(\mathrm{a})}_k(\eta,\eta_0)+\frac{1}{2}\frac{d\ln P_\delta^{(\mathrm{ad})}(\eta_0,k)}{d\eta_0}T^{(\mathrm{b})}_k(\eta,\eta_0).
 \eeq
 We used $\partial_{\eta_0}\ln \delta^{\rm ad}_\bk(\eta_0)=\frac{1}{2}\partial_{\eta_0} 
 \ln P_\delta^{(\rm ad)}(\eta_0,k)$ which is independent of $\mathcal{R}_\bk$.

\end{document}